\documentclass[12pt]{article} 

\usepackage{comment}
\usepackage{xcolor}
\usepackage{ulem}
\usepackage{cancel}
\usepackage{multirow}
\usepackage{diagbox}
\usepackage{subfigure}
\usepackage{amsfonts}
\usepackage{float}
\usepackage{epsfig}
\usepackage{mathrsfs}
\usepackage{array}
\usepackage{verbatim}
\newcommand{\PreserveBackslash}[1]{\let\temp=\\#1\let\\=\temp}
\newcolumntype{C}[1]{>{\PreserveBackslash\centering}p{#1}}

\usepackage[letterpaper]{geometry}
\def\spacingset#1{\renewcommand{\baselinestretch}%
	{#1}\small\normalsize} \spacingset{1}

\usepackage[colorlinks,bookmarksopen,bookmarksnumbered,citecolor=blue,urlcolor=blue,linkcolor=blue]{hyperref}
\usepackage{natbib}

\usepackage{amsthm,amssymb,amsmath}

\newtheorem{proposition}{Proposition}

\usepackage[ruled, linesnumbered]{algorithm2e}

\usepackage{grffile,verbatim}
\usepackage{enumitem}
\usepackage{array,multirow}

\usepackage{graphicx}
\usepackage{caption}

\usepackage{color}
\usepackage{comment}

\usepackage{prettyref,soul}
\newrefformat{corollary}{Corollary~\ref{#1}}
\newrefformat{lemma}{Lemma~\ref{#1}}
\newrefformat{theorem}{Theorem~\ref{#1}}
\newrefformat{assumption}{Assumption~\ref{#1}}
\newrefformat{condition}{Condition~\ref{#1}}

\newcommand{\argmin}{\ensuremath{\operatornamewithlimits{argmin}}}

\def\trans{^{\rm T}}

\newcommand{\bbR}{\mathbb{R}}

\newcommand{\bbL}{\mathbb{L}}

\newcommand{\sE}{\mathcal{E}}

\newcommand{\sG}{\mathcal{G}}
\newcommand{\sI}{\mathcal{I}}

\newcommand{\sR}{\mathcal{R}}
\newcommand{\sS}{\mathcal{S}}
\newcommand{\sT}{\mathcal{T}}
\newcommand{\sO}{\mathcal{O}}

\newcommand{\sV}{\mathcal{V}}
\newcommand{\sW}{\mathcal{W}}

\newcommand{\vd}{\mathbf{d}}

\newcommand{\vg}{\mathbf{g}}

\newcommand{\vi}{\mathbf{i}}

\newcommand{\vu}{\mathbf{u}}

\newcommand{\vx}{\mathbf{x}}
\newcommand{\vy}{\mathbf{y}}

\newcommand{\valpha}{\boldsymbol{\alpha}}
\newcommand{\vomega}{\boldsymbol{\omega}}
\newcommand{\vbeta}{\boldsymbol{\beta}}

\newcommand{\vpsi}{\boldsymbol{\psi}}

\newcommand{\vphi}{\boldsymbol{\phi}}

\newcommand{\vzero}{\mathbf{0}}

\newcommand{\mA}{\mathbf{A}}
\newcommand{\mB}{\mathbf{B}}

\newcommand{\mI}{\mathbf{I}}
\newcommand{\mJ}{\mathbf{J}}

\newcommand{\mL}{\mathbf{L}}

\newcommand{\mP}{\mathbf{P}}
\newcommand{\mQ}{\mathbf{Q}}
\newcommand{\mR}{\mathbf{R}}

\newcommand{\mT}{\mathbf{T}}
\newcommand{\mU}{\mathbf{U}}
\newcommand{\mV}{\mathbf{V}}
\newcommand{\mW}{\mathbf{W}}
\newcommand{\mX}{\mathbf{X}}

\newcommand{\mGamma}{\boldsymbol{\Gamma}}
\newcommand{\mOmega}{\boldsymbol{\Omega}}
\newcommand{\mLambda}{\boldsymbol{\Lambda}}

\newcommand{\mTheta}{\boldsymbol{\Theta}}
\newcommand{\vxi}{\mbox{\boldmath $\xi$}}

\newcommand{\intd}{\,\mathrm{d}}

\usepackage{relsize}

\usepackage{xr}
\makeatletter
\newcommand*{\addFileDependency}[1]{
	\typeout{(#1)}
	\@addtofilelist{#1}
	\IfFileExists{#1}{}{\typeout{No file #1.}}
}
\makeatother

\spacingset{1.5}
\newcommand{\blind}{1}

\begin{document}

\if1\blind
{
\title{Bayesian Nonlinear Tensor Regression with Functional Fused Elastic Net Prior}
\date{}
\author{Shuoli Chen\thanks{Corresponding authors.}~, Kejun He\footnotemark[1]~,
Shiyuan He, Yang Ni, and Raymond K.~W.~Wong}  
\let\svthefootnote\thefootnote 
\let\thefootnote\relax\footnotetext{Authors are listed alphabetically. Shuoli Chen (Email:~\href{mailto:csljiaj@ruc.edu.cn}{csljiaj@ruc.edu.cn}) is Student, Kejun He (Email:~\href{mailto:kejunhe@ruc.edu.cn}{kejunhe@ruc.edu.cn}) is Assistant Professor, Shiyuan He (Email:~\href{mailto:heshiyuan@ruc.edu.cn}{heshiyuan@ruc.edu.cn}) is Assistant Professor, Center for Applied Statistics, Institute of Statistics and Big Data, Renmin University of China, Beijing 100872, China. Yang Ni (Email:~\href{mailto:yni@stat.tamu.edu}{yni@stat.tamu.edu}) is Assistant Professor, Raymond K. W. Wong (Email:~\href{mailto:raywong@tamu.edu}{raywong@tamu.edu}) is Associate Professor, Department of Statistics, Texas A\&M University, College Station 77843, USA. }
\let\thefootnote\svthefootnote
\maketitle
}\fi

\begin{abstract}
Tensor regression methods have been widely used to predict a scalar response from covariates in the form of a multiway array.
In many applications,
the regions of tensor covariates used for prediction are often spatially connected with unknown shapes and discontinuous jumps on the boundaries. Moreover, the relationship between the response and the tensor covariates can be nonlinear.
In this article, we develop a nonlinear Bayesian tensor additive regression model to accommodate such spatial structure.
A functional fused elastic net prior is proposed over the additive component functions to comprehensively model the nonlinearity and spatial smoothness, detect the discontinuous jumps, and simultaneously identify the active regions.
The great flexibility and interpretability of the proposed method against the alternatives are demonstrated by a simulation study and an analysis on facial feature data. 
\end{abstract}

		
\newpage

\section{Introduction}\label{Introduction}
	
Data in the form of multiway arrays, also known as tensors, are becoming increasingly common in physical and engineering sciences. For example, \cite{yan2019structured} studied the machinability of titanium alloy where the cylinder-shaped materials are represented by multidimensional arrays. \cite{yue2020tensor} performed quality inspections of nanomanufacturing processes with Raman spectral imaging data which are formulated as a tensor. 
\cite{zhong2022image} proposed a tensor-based approach to handle the spatial and temporal structures of image outputs in the automatic control processes of semiconductor manufacturing.
In hot rolling processes, multiple sensors record the temperature, current, torque, speed at an equal time interval, generating multiple signals in form of tensors \citep{miao2021structural}. \cite{shi2023process} provided a good review for some recent applications of  statistical tensor methods in manufacturing quality improvement. 
Tensor data are also important in many other areas such as chemometrics \citep{andersen2003practical}, text mining \citep{chew2007cross}, and recommendation systems \citep{park2009pairwise}. 
Among the successful applications of tensor data analysis, using tensor regression to decode the relationship between a scalar response and the covariates of a tensor structure has attracted considerable attentions.  
In condition monitoring and industrial asset management, \cite{fang2019image} applied a tensor regression model to predict the residual lifetime of a rotating machinery according to the degradation image streams acquired using an infrared camera. 
In neuroscience, researchers apply tensor regression methods to predict diseases and disorders such as Alzheimer’s disease \citep{kandel2013predicting} and autism spectrum disorder \citep{ecker2013intrinsic} based on the magnetic resonance imaging or diffusion tensor imaging of human brain. 
	
A general scalar-on-tensor regression model between a $D$-way tensor of covariates $\mX\in \mathbb{R}^{P_1\times\cdots\times P_D}$ and a response $Y\in\mathbb{R}$ can be formulated via a regression function $f: \mathbb{R}^{P_1\times\cdots\times P_D}\rightarrow \mathbb{R}$ and an additive noise: $Y = f(\mX) + \epsilon$. 
The majority of existing tensor regression methods adopts the linear regression form $f(\mX) = \sum_{i_1,\cdots,i_D}X_{i_1,\cdots,i_D} \beta_{i_1,\cdots,i_D}$ where $\beta_{i_1,\cdots,i_D}$ is the $(i_1,\cdots,i_D)$-th element of the tensor coefficient $\vbeta\in R^{P_1\times\cdots\times P_D}$ to be estimated. 
To overcome the difficulty of estimating a huge number of coefficients in many tensor applications, \citet{zhou2013tensor} proposed a linear tensor regression model with a low-rank structure of $\vbeta$ via the CANDECOMP/PARAFAC (CP) decomposition \citep{harshman1970foundations}.
Additional regularization methods such as the lasso \citep{tibshirani1996regression} and the ridge \citep{hoerl1970ridge} were also suggested 
to  obtain a consistent and interpretable estimator. 
\citet{BTR2017} proposed Bayesian Tensor Regression (BTR), which again utilized the CP decomposition. With carefully constructed shrinkage priors, BTR is able to shrink parameters at both local and global levels, and select the rank automatically.
Some other works of tensor linear regression are based on different types of decomposition including Tucker decomposition \citep{tucker1966some} on the coefficient tensor $\vbeta$ \citep{li2018tucker}. 
	
However, the assumption that the tensor covariates can predict the response through a linear regression function is too restrictive and can be violated in many applications.
For instance, in the field of financial analysis, \cite{li2016tensor} found that the nonlinearity exists in the relationship between stock movements and information sources in the form of tensor data.
To model the nonlinearity of regression function $f$ while keeping the inherent structural information of the original tensor, 
\cite{zhao2013kernelization, zhao2014tensor} placed a Gaussian process prior over the regression function where the covariance function is a product kernel based on the unfoldings of tensor covariates. 
With a rank-$1$ CP decomposition $\mX= \vx_{1}\circ\cdots\circ \vx_{D}$ where $\circ$ denotes the outer product and $\vx_d$ is a $P_d$-dimensional vector, \cite{signoretto2013learning} and \cite{kanagawa2016gaussian} considered a regression model $f(\mX) = \sum_{r=1}^R\prod_{d=1}^D f_r^{(d)}(\vx_{d})$ with a Gaussian process prior over each $f_r^{(d)}$, $d=1,\dots, D$. 
Extending the rank-$1$ assumption, a more flexible model $f(\mX) = \sum_{r=1}^R\sum_{m=1}^M\prod_{d=1}^D f_r^{(d)}(\vx^{(m)}_{d})$ with $\mX= \sum_{m=1}^M\vx^{(m)}_{1}\circ\cdots\circ \vx^{(m)}_{D}$ was proposed in \cite{imaizumi2016doubly}. 
Unfortunately, a number of multi-dimensional functions have to be estimated in the above work, which will suffer from the curse of dimensionality when some $P_d$'s are large.
An alternative approach of modeling the nonlinear regression function is using the similar idea of additive models \citep{stone1985additive} on the vector of covariates. 
Nonparametric additive models have recently been extended to tensor covariates with elastic net \citep{zhou2020broadcasted} and the group lasso penalty \citep{hao2021sparse}. They again exploit the tensor structure through CP decomposition of the tensor coefficient. 

In many applications, the tensor of covariates (e.g., a 3D image) is a collection of observations at a regular grid over a multidimensional continuous domain.
One common observation in the corresponding applications is the existence of spatially contiguous active regions with unknown shapes and discontinuous jumps on the boundaries of regions, especially in image data. 
For example, in neuroscience, the pathological studies show that the brain voxels that have significant effects to the diseases are expected to be sparse and organized into several spatially connected regions \citep{michel2011total, fiot2014longitudinal}. 
Therefore, the presence of multiple piecewise smooth regions should be considered in the regression function $f$. 
Although there exist prior works that are related to the modeling of this spatial structure, such as \citet{xin2014efficient, goldsmith2014smooth, li2015spatial, wang2017generalized, beer2019incorporating}, most make the linear assumption on the regression function.
One notable exception is \cite{marx2011multidimensional},
which proposed a nonlinear tensor regression with spatial similarity through a single-index model.
However, their method does not produce sparse estimation, and thus the important subregions are hard to be identified using their model. 
In this work, we propose a novel Bayesian tensor additive regression model that incorporates the spatial structure of tensor covariates and strikes a good balance between flexibility and interpretability. 
More precisely, we design a prior called functional fused elastic net (FEN) over the nonlinear additive component functions to adaptively learn
the spatial smoothness of the component functions within
unknown connected regions. 
The spatial smoothness is achieved by the graph Laplacian of the adjacent entries, and discontinuous jumps between distinct regions are detected by the $\ell_1$ fusion of the adjacent entries. 
With spline representation, we apply the idea of the thresholding method \citep{ni2019bayesian, BayesianNetworkMarker2020} on the coefficients to achieve sparsity and identify the important regions.
A crucial advantage of thresholding method against the common alternatives, such as spike-and-slab priors \citep{mitchell1988bayesian} and Bayesian credible intervals \citep{chen1999monte}, is its low computation cost and the ability to drop the inactive signals without increasing the predictive error.
The posterior inference is carried out through a Markov chain Monte Carlo (MCMC) method with the Metropolis-adjusted Langevin Algorithm \citep[MALA,][]{roberts1998optimal}. 
To the best of our knowledge, our work is the first to integrate the spatial smoothness and discontinuous jumps for sparse nonlinear tensor regression. 
	
The rest of this paper is organized as follows. 
In Section \ref{Tensor_Additive}, we present the tensor additive model and introduce the spatially piecewise smooth structure to integrate the idea of sparsity, spatial smoothness, and discontinuous jumps. 
Section \ref{Model_Priors} proposes the functional FEN prior for the component functions of the tensor additive model and illustrates its properties with some examples. 
Using spline expansion to approximate each additive component function, a Bayesian hierarchical model is formulated on the spline coefficients, and a posterior sampling algorithm is described. 
A simulation study and a real application on facial feature data are respectively presented in Sections \ref{simulation} and \ref{real_application} to demonstrate the advantages of the proposed model over existing alternatives. 
We finally summarize this article in Section \ref{sec:dis} with some concluding remarks.
	
\section{Tensor Additive Regression Model}\label{Tensor_Additive}
	
We consider the scalar-on-tensor regression setting where the covariate $\mX\in \bbR^{P_1\times\cdots\times P_D}$ is a $D$-way tensor of dimension $P_1\times\cdots\times P_D$ and the response $Y\in\bbR$ is a scalar. 
The element $X_{\vi}$ of $\mX$ is indexed by  $\vi \in\sI = \{(i_1, i_2, \cdots, i_D):   1\leq i_d \leq P_d, 1\leq d\leq D\}$. Without loss of generality, we assume $X_{\vi}\in [0,1]$ for all $\vi$. 
The number of elements in $\mX$ can be much larger than the sample size 
in many applications. 
For example, the magnetic resonance imaging dataset considered in \cite{zhou2013tensor} consists of $776$ patients with the number of covariates up to $256\times 198\times 256 = 12,976,128$.
High dimensionality leads to significant difficulties in modeling the nonlinear regression function.
A natural nonlinear regression model is
a tensor additive model:
\begin{equation}\label{eqn:model}
		Y = \mu + \sum_{\vi\in\sI} f_{\vi}(X_{\vi}) + \epsilon, \quad\epsilon\sim N(0, \sigma_{\epsilon}^2),
\end{equation}
where $f_{\vi}$'s are nonlinear functions such that $\int_0^1 f_{\vi}(x)\intd x = 0$ for all $\vi$ (for identifiability purposes).
However, even with the additive model assumption, there are still a potentially huge number of univariate nonparametric functions to be estimated.
With a limited amount of data, it is often challenging to estimate these functions well.
Furthermore, there are three types of useful structures in tensor regressions, which are not incorporated by model \eqref{eqn:model}. 
\begin{figure}[h!]
	\centering
	\includegraphics[width=\textwidth]{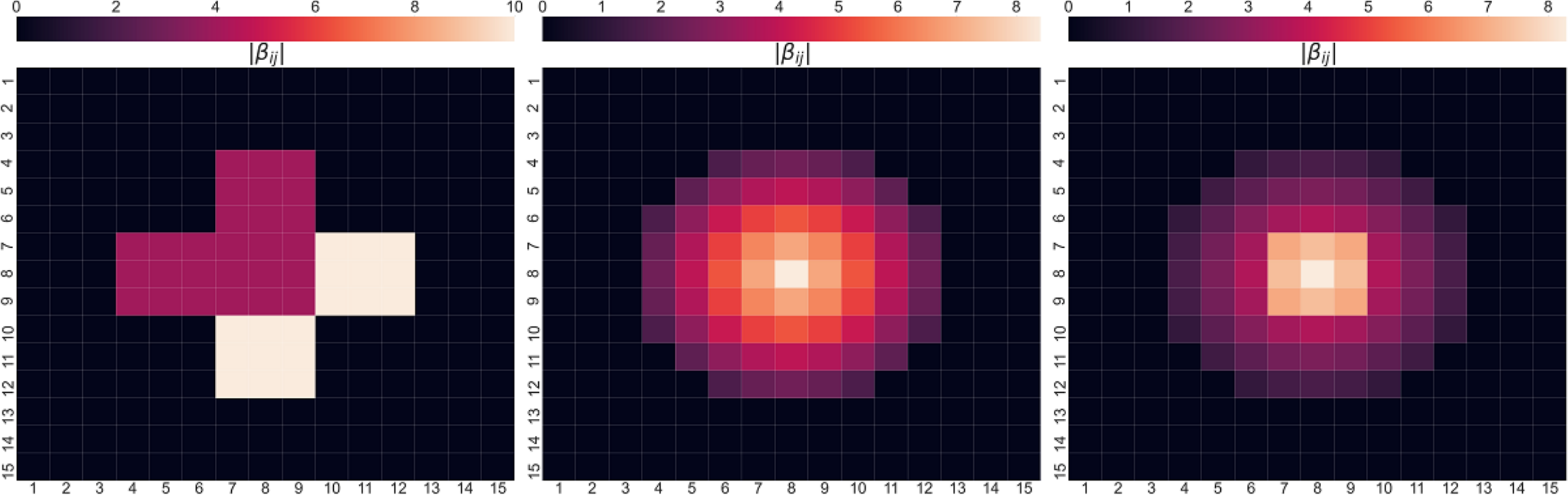}
	\caption{Three examples of $2$-way tensor additive model with spatially piecewise smooth structure. The heatmaps show the magnitude of each additive component.}
	\label{fig:spatialsmooth}
\end{figure}
	
\textbf{Sparsity}. In many real applications, only a few entries of the tensor covariates may be relevant to predict the response. Take neuroimaging as an example, the brain is believed to have dedicated regions for different tasks. For instances, the visual cortex in human brains controls visual functions \citep{grill2004human} and the frontal lobe is responsible for reasoning \citep{collins2012reasoning}.
We thus generally expect many additive component functions in \eqref{eqn:model} to be zero (i.e., $f_{\vi}\equiv 0$) for predicting reasoning and visual-related outcomes. In the following, the sets of $\vi$ where the additive component function $f_{\vi}$ is non-zero and zero are called active regions and non-active regions, respectively.
	
	\textbf{Spatial smoothness}. We further assume  the additive model~\eqref{eqn:model} to be endowed with a \textit{spatially smooth functional structure}, which means that the  functions $f_{\vi}$'s vary smoothly with respect to the location index $\vi$. Specifically, the functions $f_{\vi}$'s are spatially smooth with respect to a graph 
	$\sG = (\sI, \sE)$,
	where $\sE$ is the neighboring relationship set for the location index set $\sI$. A pair of indices $(\vi,\vi')\in\sE$ are connected by an edge when $X_{\vi}$ and  $X_{\vi'}$ are neighboring elements in the tensor of covariates $\mX$. Equivalently, $(\vi, \vi')\in \sE$ if $\Vert \vi -\vi'\Vert_1 = 1$, where $\|\cdot\|_1$ represents the $\ell_1$-norm.
	For the additive model~\eqref{eqn:model} to be spatially smooth with respect to $\sG$,  functions $f_{\vi}, f_{\vi'}$ with $(\vi,\vi')\in\sE$ are likely to be similar to each other.
	
	\textbf{Discontinuous jumps}. 
	Sparsity and spatial smoothness together require the functions to be smoothly decaying to zero towards the boundary of an active/non-zero region. This may not be realistic. In a natural image or neuroimage, a pixel (or voxel) at the boundary of an active region could have significant effect on the response.
	Our work aims to address this challenging issue, by developing a spatially smooth model that allows for occasional discontinuous jumps, i.e., if supported by data, a few $f_{\vi}$'s can vary non-smoothly from its neighbors.
	
	Combining spatial smoothness and discontinuity, we obtain
	a hybrid structure, which we call \textit{spatially piecewise smooth functional structure}. 
	More specifically, in this structure, the index set $\sI$ can be divided into a few distinct spatially connected regions $\sI_{1}, \cdots, \sI_{C}$, and the component functions within the same region are spatially smooth. Discontinuity are allowed on the boundary between regions.
	
	Figure~\ref{fig:spatialsmooth} illustrates the various types of spatially piecewise smooth functional structures that our model can handle. It shows the heatmap of $|\beta_{\vi}|$ for  function 
	$f_{\vi}(X_{\vi}) = \beta_{\vi} X_{\vi}$, which is linear for the simplicity of illustration. In the left panel, the active (non-black) regions can be divided into three pieces. Inside each piece,  $|\beta_{\vi}|$ is spatially smooth (in fact, it is a constant).  There are discontinuity jumps between the active and non-active regions and between each pair of active regions. The middle panel  simply contains one active region and is overall smooth. The right panel has a discontinuity jump at the central square, and is spatially smooth within the central square and the surrounding circle, respectively.

\section{Bayesian Model}\label{Model_Priors}
In this section, we develop a Bayesian hierarchical model for the inference of the tensor additive model~\eqref{eqn:model}. We propose a \textit{functional fused elastic net (functional FEN)} prior to deal with the spatially piecewise smooth functional structure and illustrate its advantage through two simple numerical experiments. 
Using basis representation, we show that the proposed functional FEN prior can be transferred to a proper prior on the the basis coefficients. 
An efficient computational algorithm for the posterior inference is also developed.

\subsection{Functional Fused Elastic Net Prior}\label{FFEN_section}
To construct a prior distribution that encourages sparsity, each $f_{\vi}$ is parameterized as the product of a latent function $g_{\vi}\in C^2[0, 1]$ and a hard thresholding function $\mathbf{1}_{\{\Vert g_{\vi}\Vert_{\bbL_2}^2 > \lambda \}}$, i.e., \begin{equation}\label{eqn:f_g_threshold}
    f_{\vi} = g_{\vi}\cdot\mathbf{1}_{\{\Vert g_{\vi}\Vert_{\bbL_2}^2 > \lambda \}}, \quad \vi \in \sI, 
\end{equation} 
where $\lambda$ is the thresholding parameter.
Roughly speaking, $f_{\vi}$ is thresholded to exact zero $f_{\vi} \equiv 0$ whenever the latent function $g_{\vi}$ has a small magnitude.
Using the form of \eqref{eqn:f_g_threshold}, the spatially piecewise smooth functional structure on $f_{\vi}$ can be equivalently modeled on $g_{\vi}$.

Let $\ell[0, 1]$ denote the set of affine functions on the interval $[0, 1]$, i.e., 
\begin{equation}\label{eqn:defAffine}
    \ell[0, 1]=\{l(x): l(x)=a+bx\}.
\end{equation}
Denote the projection operator from the space of the second order Sobolev space $\sW^2_2[0, 1]$ onto $\ell[0, 1]$ by $\mathcal{P}$.
We propose a functional FEN prior distribution
for the set of all the latent functions $\boldsymbol{G} = \{g_{\vi}(x):\ \vi\in\sI \}$:
\begin{align} \label{eqn:ffenp}
	p\big(\boldsymbol{G}|& \,\delta, r_1, r_2\big) \propto 
	\exp\Big\{- \delta \sum_{\vi \in\sI} \sR(g_{\vi})
	- r_1\sum_{(\vi,\vi') \in \sE}\Vert g_{\vi} - g_{\vi'}\Vert_{\bbL_2} -r_2\sum_{(\vi,\vi') \in \sE}\Vert g_{\vi} - g_{\vi'}\Vert_{\bbL_2}^2\Big\}, 
\end{align}
where $\sR(g_{\vi}) =\Vert g''_{\vi}\Vert_{\bbL_2}^2 +\delta'\| \mathcal{P} g_{\vi}\|_{\bbL_2}^2$ measures the roughness of $g_{\vi}$ with $g''_{\vi}$ being the second derivative of $g_{\vi}$ and $\delta' \in \mathbb{R}^+$. 
The second summation in the prior distribution~\eqref{eqn:ffenp} is the functional \textit{fusion} term, which encourages local constant structure and helps build the piecewise structure. The third summation is the functional \textit{Laplacian} term, which encourages spatial smoothness.

The fusion and the Laplacian terms of the functional FEN prior distribution~\eqref{eqn:ffenp} can be viewed as an \textit{adaptive} Laplacian prior distribution. To see this, we use a Gaussian  scale mixture identity as follows:
$$e^{- b| x| }  = \int \frac{1}{\sqrt{\pi \omega}} \exp\bigg(
\frac{b^2 x^2}{4\omega} \bigg) \cdot e^{-\omega}   \intd \omega.$$
We can rewrite the fusion term in \eqref{eqn:ffenp} through  independent latent random variables $\omega_{\vi\vi'}$, $(\vi,\vi')\in\sE$, following the standard exponential distribution, 
\begin{align} 
		p\big(\boldsymbol{G}| \,\delta,r_1, r_2, \omega_{\vi\vi'} \big) &\propto
		\prod_{(\vi,\vi')\in\sE}	\frac{1}{\sqrt{\omega_{\vi\vi'}} }
		\exp\Big\{- \delta\sum_{\vi \in\sI}
		\sR(g_{\vi}) -\sum_{(\vi,\vi') \in \sE}
		\big(r_2+\frac{r_1^2}{4\omega_{\vi\vi'} }\big)
		\Vert g_{\vi} - g_{\vi'}\Vert_{\bbL_2}^2\Big\}, 
		\nonumber \\
		\omega_{\vi\vi'} &\stackrel{\rm i.i.d.}{\sim} \mathrm{Exp}(1) \text{ for all } (\vi,\vi')\in\sE. \nonumber 
\end{align}
Using this representation, the second and the third summation in~\eqref{eqn:ffenp} are merged into a single term.
The prior distribution generally encourages the neighboring functions to be similar, i.e., with small $\bbL_2$ distance.
When $\omega_{\vi\vi'}$ is close to zero, its contribution to the prior distribution could be very large. 
Thus, for the corresponding neighboring functions $g_{\vi}$ and $g_{\vi'}$, the prior has the tendency to push them towards being identical. In the next subsection, we discuss more properties of the functional FEN prior and show how the fusion and the Laplacian terms successfully accommodate a spatially piecewise smooth structure.

\subsection{Properties of the Fuison and Laplacian Prior}\label{motivation}
For the proposed functional FEN prior distribution~\eqref{eqn:ffenp}, both the functional fusion term and the functional Laplacian term play indispensable roles. 
For simplicity, we illustrate their importance  via a special setting where each additive component function is linear with $f_{\vi}(X_{\vi}) = X_{\vi}\beta_{\vi}$ and $\beta_{\vi}\in\mathbb{R}$, $\vi \in \sI$. 
In this setting, the functional FEN prior~\eqref{eqn:ffenp} reduces to a prior on the scalars $\beta_{\vi}$'s as
\begin{align} \label{eqn:prior_scalar}
	p(\vbeta|\delta,r_1,r_2) \propto \exp\Big\{
	-\delta \sum_{\vi \in\sI} \beta_{\vi}^2
	- r_1\sum_{(\vi,\vi') \in \sE}\vert \beta_{\vi} - \beta_{\vi'}\vert -r_2\sum_{(\vi,\vi') \in \sE}(\beta_{\vi} - \beta_{\vi'})^2\Big\}.	
 \end{align}
From \eqref{eqn:prior_scalar}, we observe that when $\delta = r_2= 0$, the corresponding prior of $p(\vbeta|0,r_1,0)$ degenerates to the generalized fused lasso \citep{Tibshirani2005SparsityAS}.
When $\delta = r_1 = 0$, the FEN prior reduces to Laplacian prior or Gaussian Markov random field \citep{rue2005gaussian}.
When $\delta = 0$, the corresponding (negative log) FEN prior, i.e., $-\log p(\vbeta|0,r_1,r_2)$, is equivalent to the graph-fused elastic net penalty \citep{tec2019large}.

\begin{figure}[h!] 
	\centering 
	\includegraphics[width=\textwidth]{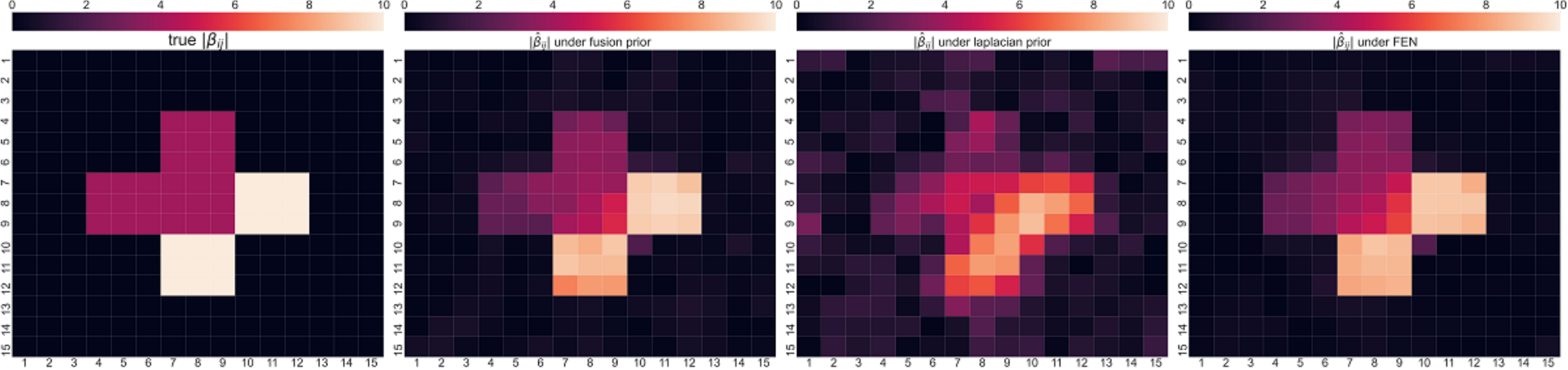} 
	\caption{A toy simulation where the component functions are linear, i.e. $f_{\vi}(X_{\vi}) = X_{\vi}\beta_{\vi}$. From left to right, the four panels correspond to  the true values of $\beta_{\vi}$, the posterior mean of  $\beta_{\vi}$'s with the fusion prior, the Laplacian prior and FEN prior, respectively.}
	\label{toy_expr}
\end{figure}

\begin{figure}[h!] 
	\centering 
	\includegraphics[width=\textwidth]{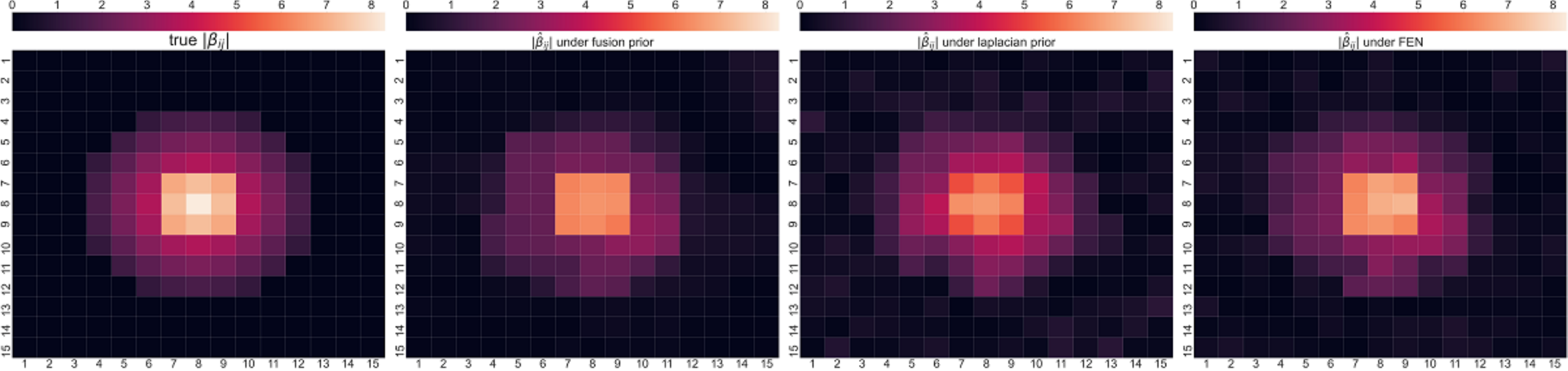} 
	\caption{A toy simulation where the component functions are linear, i.e. $f_{\vi}(X_{\vi}) = X_{\vi}\beta_{\vi}$. From left to right, the four panels correspond to  the true values of $\beta_{\vi}$, the posterior mean of  $\beta_{\vi}$'s with the fusion prior, the Laplacian prior and FEN prior, respectively.}
	\label{toy_expr3}
\end{figure} 	
We conduct two simple experiments to illustrate the properties of fusion prior and Laplacian prior, and show the performance gain of FEN by combining them.
For these two experiments, we set $\sI = \{(i,j): 1\leq i,j\leq 15 \}$, and the matrix covariates are generated by $X_{ij}\stackrel{\mathrm{i.i.d.}}{\sim} \text{Unif}(0,1)$.
The responses are generated according to $y = \sum_{1\leq i,j\leq 15} X_{ij}\beta_{ij} + \epsilon$ with $\epsilon\sim N(0,1)$, and the true values of $\beta_{ij}$'s are shown in the leftmost panels of Figures~\ref{toy_expr} and \ref{toy_expr3}, respectively for two experiments. These settings correspondingly feature the true model with the following structures:
1) a spatially piecewise  constant structure, 
and 
2) a spatially piecewise smooth structure. We generate $N=100$ observations for each setting and repeat the experiments for $30$ times. The posterior distribution of $\vbeta$ is given by
\begin{align} \label{eqn:top_model} 
	p(\vbeta |D_N,\delta, r_1,r_2 ) &\propto \exp\Big\{-\frac{1}{2} \sum_{n=1}^{N} (y_n - \langle \vbeta, \mX^{(n)}\rangle
	)^2  \Big\} \times p(\vbeta|\delta,r_1,r_2), 
 \end{align}
where the prior distribution $p(\vbeta|\delta,r_1,r_2)$ is given in \eqref{eqn:prior_scalar}. For simplicity, we fix $\delta = 0$ and vary the hyperparameters $r_1, r_2$ 
to achieve the fusion prior ($r_2 = 0$), the Laplacian prior ($r_1 = 0$), and the general FEN prior. 
For the FEN prior we adopt parameterization $r_1 = r/\rho$, $r_2 = (1 - r)/\rho$ with candidate grids $r \in \{1, 0.75, 0.5, 0.25, 0\}$ and $\rho \in \{0.3, 0.6, 1.2, 2.4, 4.8\}$. 
MALA \citep{roberts1998optimal} is applied to draw posterior samples from the model. The hyperparameters are selected as those with best predictive performance on a validation dataset.
We randomly pick one replication from each experiment setting and show the posterior mean of $\vbeta$ from the fusion, Laplacian, FEN priors in the second, third, and fourth panels of Figures \ref{toy_expr} and \ref{toy_expr3}. 
The performances of methods are also evaluated in terms of $\text{MSE} = \frac{1}{15\times 15}\sum_{i,j}(\beta_{ij}-\widehat\beta_{ij})^2$ where $\widehat\beta_{ij}$ is the posterior mean of $\beta_{ij}$. The average MSE over 30 replicates are summarized in Table~\ref{table_toy}.

\begin{table}[t]
	\centering
	\caption{The performance of fusion, Laplacian and FEN priors under  different true models for 30 random replicates. The numbers in the parentheses are the standard errors.}
	\resizebox{\textwidth}{!}{
		\begin{tabular}{ c| c c c |  c c c}
			\hline
			True Model & \multicolumn{3}{c|}{spatially piecewise constant} & \multicolumn{3}{c}{spatially piecewise smooth}\\
			\hline
			Prior & fusion & Laplacian  & FEN & fusion & Laplacian  & FEN\\
				\hline 
				\hline 
			MSE & $0.4286$ $(0.0204)$ &$2.4860$ $(0.0635)$  &$ 0.4317$ $(0.0189)$ &$0.1707$ $(0.0080)$ &$0.3265$ $(0.0075)$ &$0.1589$ $(0.0059)$ \\
				\hline
			\end{tabular}
		}
	\label{table_toy}
\end{table}

Figure~\ref{toy_expr} and Table \ref{table_toy} reveal that, when the true model has a spatially piecewise constant structure, the fusion prior ($r_2=0$) has a smaller MSE than the Laplacian prior ($r_1=0$).  The estimated  $\beta_{ij}$'s from the fusion prior (the second panel) recover the true signal pattern reasonably well. However, the true pattern has been smoothed out by the Laplacian prior (the third panel). 
The FEN prior selects $r_2 = 0$ in all $30$ replicates, and hence its performance (the fourth panel) is similar to that of the fusion prior.

The above results demonstrate the advantage of fusion prior over Laplacian prior in estimating spatially piecewise constant model, which is consistent with the findings in \cite{Tibshirani2005SparsityAS} and \cite{ little2010sparse}. 
However, the fusion prior tends to force similar neighboring values to be identical, and so it may introduce biases when the true values are not exactly constant.
Figure~\ref{toy_expr3} shows an example the Laplacian prior and the fusion prior can be combined to tackle more challenging settings. As shown in the leftmost panel, the true model is spatially piecewise smooth. There are discontinuity jumps on the boundary between a center square piece and a surrounding circle piece, and the signals vary smoothly within each piece.
Neither the fusion prior nor the Laplacian prior estimates $\beta_{ij}$'s accurately in this case.
The fusion prior over-shrinks the coefficient in the center square piece and the surrounding circle piece to a constant, while the Laplacian prior over-smooths the estimates globally. By contrast, the FEN prior, which combines the fusion and the Laplacian priors, is able to capture the corresponding piecewise smooth structure and has the lowest MSE.

\subsection{Spline Representation of Functions}\label{Spl_Repre}
To facilitate the estimation of the unknown functions, we expand $g_{\vi}(x) = \sum_{k=1}^K \alpha_{\vi k} \phi_k (x)$ via a vector of spline basis functions $\vphi(x) = (\phi_1(x), \dots, \phi_K(x))\trans $, where $\valpha_{\vi} = (\alpha_{\vi 1}, \dots, \alpha_{\vi K})\trans $ is the vector of spline coefficients, $\vi \in \sI$. Denote $\valpha\trans = (\valpha_{\vi}\trans)_{\vi \in \sI} \in \mathbb{R}^{P_1\times\cdots\times P_D\times K}$.
We require the vector of basis functions $\vphi(\cdot)$ to have the following properties.
\begin{enumerate}[nosep]
	\item[(i)] The basis functions are centered, i.e., $\int \vphi(x) \intd x =\vzero$. This guarantees $\int g_{\vi}(x) \intd x = \int\valpha_{\vi} \trans\vphi (x)\intd x = 0$ for any $\valpha_{\vi}\in\bbR^K$ and thus $\int f_{\vi}(x) \intd x = 0$ due to \eqref{eqn:f_g_threshold}. 	
    \item[(ii)] The basis functions are orthonormal, i.e., $\int \vphi(x)\vphi(x)\trans \intd x = \mI_K$.
	As such, the $\bbL_2$ norm of function $g_{\vi}$ can be directly evaluated as the Euclidean norm  of the spline coefficients,  $\Vert g_{\vi}\Vert_{\bbL_2}^2 = \Vert\valpha_{\vi}\Vert_2^2$. This also facilitates the representation of \eqref{eqn:f_g_threshold}
	as $f_{\vi}(X_{\vi}) = \vphi(X_{\vi})\trans \vbeta_{\vi}$ with 
	\begin{equation} \label{sparse_dec}
		\vbeta_{\vi} = \valpha_{\vi}\cdot\mathbf{1}_{\{\Vert \valpha_{\vi}\Vert_2^2 > \lambda \}}.	
    \end{equation}
	\item[(iii)] The second derivatives of the basis functions are orthogonal, i.e., $\mOmega := \int \vphi''(x)\vphi''(x)\trans \intd x$ $= \mathrm{diag}(\omega_{11},\omega_{22}, \cdots,\omega_{KK})$. The $\bbL_2$ norm of  $g''_{\vi}$ can thus be directly evaluated by the weighted Euclidean norm  of the spline coefficients, i.e., $\Vert g_{\vi}''\Vert_{\bbL_2}^2 = \sum_{k=1}^K \omega_{kk} \alpha_{\vi k}^2$.
\end{enumerate}
In the above, the first property is for the identifiability of the additive model \eqref{eqn:model}. 
The second and the third reduce the complexity of calculating $\bbL_2$ norm of $g_{\vi}$ and $g_{\vi}''$ from $\sO(K^2)$ to $\sO(K)$. A vector of basis functions $\vphi$ that satisfies these conditions can be constructed from B-spline basis functions, and the details are provided in Section~\ref{supp:app:spline} of the Appendix. We also show that the first basis function, $\phi_1$, in the constructed bases satisfies $\phi_1 \in \ell [0, 1]$ as defined in \eqref{eqn:defAffine}.

Using the basis functions $\vphi$, the functional FEN prior~\eqref{eqn:ffenp} can be written as
\begin{align}\label{eqn:modelprior}
	p(\valpha|\delta, r_1,r_2) \propto \exp\Big(- \delta\sum_{\vi \in\sI} \valpha_{\vi}\trans\mR \valpha_{\vi}    - r_1\sum_{(\vi,\vi') \in \sE}\Vert \valpha_{\vi}-\valpha_{\vi'}\Vert_2 - r_2\sum_{(\vi,\vi') \in \sE}\Vert\valpha_{\vi}-\valpha_{\vi'}\Vert_2^2\Big).
\end{align}
In \eqref{eqn:modelprior}, $\mR = \mOmega + \delta'\Vert\phi_1\Vert_{\bbL_2}^2\mathbf{e}_1\mathbf{e}_1\trans$, where $\mathbf{e}_1 = (1, 0, \dots, 0)\trans$.

\subsection{The Hierarchical Bayesian Model}\label{Baysian_model}
	
We now summarize our hierarchical model. Given the intercept $\mu$, the spline coefficients $\valpha$, the residual variance $\sigma^2$, and the thresholding parameter $\lambda$, the response $y_n$ for the $n$-th observation follows a Gaussian distribution,
\begin{align} \label{eqen:likelihood}
	y_n|\mu, \valpha, \mu, \sigma^2, \lambda\stackrel{\mathrm{i.i.d.}}{\sim} N\Big(\mu+\sum_{\vi\in\sI}\phi(X^{(n)}_{\vi})\trans \valpha_{\vi}\cdot \mathbf{1}_{\{\Vert\valpha_{\vi}\Vert_2^2 > \lambda\}}, \sigma^2\Big).	
\end{align}
A weakly informative Gaussian prior is imposed for $\mu$, and a generalized inverse Gaussian distribution $\text{GIG}(p, a, b)$ is imposed for the thresholding parameter $\lambda$, i.e.,
\begin{align}\label{prior_mu_lambda}
	\mu\sim N(0, \sigma_{\mu}^2)\quad \mbox{and} \quad \ln p(\lambda)\propto (p-1)\ln\lambda -\frac{a/\lambda + b\lambda}{2}.
 \end{align}
The generalized inverse Gaussian distribution keeps $\lambda$ away from $0$ and meanwhile prevents $\lambda$ from being too large.
The prior \eqref{eqn:modelprior} of the spline coefficients $\valpha$ is re-parameterized as
\begin{align}\label{naive_p} 
	&\hspace{-9.75pt} p(\valpha\mid \delta, r, \sigma^2, \rho_{\valpha}) = \nonumber\\
	&\hspace{-9.75pt} \frac{1}{C_{\delta, \sigma^2}}
		\cdot\exp\Big(-\frac{\delta\sum_{\vi\in\sI}\valpha_{\vi}\trans \mR\valpha_{\vi}}{\sigma^2}- \frac{r\sum_{(\vi,\vi') \in \sE}\Vert\valpha_{\vi}-\valpha_{\vi'}\Vert_2^2 + (1-r)\sum_{(\vi,\vi') \in \sE}\Vert\valpha_{\vi}-\valpha_{\vi'}\Vert_2}{2 \sigma^2\rho_{\valpha}}\Big),
\end{align} 
where $C_{\delta, \sigma^2}$ is a normalizing term, $\delta$ and $\rho_{\valpha}$  control the informativeness of the prior, and $r$ controls the relative weights of the Laplacian prior and fusion prior. 
The normalizing term $C_{\delta, \sigma^2}$ depends on $\delta$ and $\sigma^2$  and is not analytically available. Therefore, to facilitate computation, we propose a joint prior for $\sigma^2$ and $\delta$,
\begin{align}\label{prior_del_sig}
	p(\delta,\sigma^2) \propto C_{\delta, \sigma^2}\cdot\delta^{p_0 - 1}\exp(-\delta)\cdot\Big(\frac{1}{\sigma^2}\Big)^{p_1+1}\exp\Big(-\frac{1}{\sigma^2}\Big), \end{align}
which includes the normalizing term $C_{\delta, \sigma^2}$ in \eqref{naive_p}. This construction allows the normalizing term to be canceled out when deriving the full conditional of $\delta$ and $\sigma^2$.
	 
However, special care is needed to ensure that \eqref{prior_del_sig} is proper and also weakly informative. 
For propriety, the integral of~\eqref{prior_del_sig} is finite if and only if the integral with respect to $\delta$ in the neighborhood of $0$ and the integral with respect to  $\sigma^2$ in the neighborhood of $+\infty$  are both finite. Hence we need to derive the order of magnitude of $ C_{\delta, \sigma^2}$ as $\sigma^2\to\infty$ and $\delta\to 0$. Proposition~\ref{cons_thm} below shows $p_0$ and $p_1$ in \eqref{prior_del_sig} should be at least larger than $K/2$ and $P_1\cdots P_DK/2$, respectively.
	
\begin{proposition}\label{cons_thm}
	The order of magnitude of the normalizing term $C_{\delta,\sigma^2}$ satisfies: 
	(i) with respect to $\delta$, $C_{\delta,\sigma^2}$ is of order $({1}/{\delta})^{K/2}$ as $\delta\to0$, and of order $({1}/{\delta})^{P_1\cdots P_DK/2}$ as $\delta\to+\infty$;
	(ii) with respect to $\sigma^2$, $C_{\delta,\sigma^2}$ is of order $(\sigma^2)^{(2P_1\cdots P_D - 1)K/2}$ as $\sigma^2\to 0$, and of order $(\sigma^2)^{P_1\cdots P_DK/2}$ as $\sigma^2\to+\infty$.
	\end{proposition}
	The proof is provided in Section~\ref{proper_hyp} of the Appendix.
	For weak informativeness, we suggest to standardize the response variable so that the variance $\sigma^2$ of noise $\epsilon$ should concentrate on $[0,1]$.
Because Proposition \ref{cons_thm} shows that $C_{\delta,\sigma^2}$ is of order $\big(\sigma^2\big)^{(2P_1\cdots P_D - 1)K/2}$ as $\sigma^2\to 0$, we set $p_1=(2P_1\cdots P_D - 1)K/2$  to balance the magnitude of $C_{\delta,\sigma^2}$ and thus make \eqref{prior_del_sig} weakly informative with respect to $\sigma^2$ (similar to an inverse-gamma prior with the shape parameter close to 0 near the origin). 
As for the hyperparameter $p_0$, it is associated with the parameter $\delta$, which controls the smoothness of the function. We will determine $p_0$ in a data-adaptive way and present the details in Section~\ref{supp:validation} of the Appendix.

Although the approximation of additive component functions $f_{\vi}$'s and the definition of $\mR$ in prior \eqref{naive_p} are based on the vector of spline basis functions $\vphi$,
the posterior distribution of $\sum_{\vi\in\sI}f_{\vi}(x)$ remains unchanged for the proposed hierarchical model 
if an equivalent vector of orthonormal bases of the same spline space is employed.
This invariant property of our hierarchical model is summarized in Proposition~\ref{prop:priorinvariant} and its proof is presented in Section~\ref{supp:app:invarant} of the Appendix.

\begin{proposition} \label{prop:priorinvariant}
The inference for tensor additive regression~\eqref{eqn:model} is invariant with respect to an orthonormal transformation of the basis functions. That is,
for $\vphi_{\mQ} = \mQ\vphi$ where $\mQ \in \bbR^{K\times K}$ is orthonormal, the posterior distribution of $f := \sum_{\vi\in\sI}f_{\vi}$ remains unchanged. 
\end{proposition}

\subsection{Posterior Sampling Algorithm}\label{MALA_Warmstart}
We apply a hybrid MCMC method to obtain the posterior samples of $\{\mu, \valpha, \lambda, \sigma^2, \delta\}$ from the hierarchical model \eqref{eqen:likelihood}--\eqref{prior_del_sig}. 
In particular, the MALA \citep{roberts1998optimal} is used to sample $\mu$ and $\valpha$; the parameters $\sigma^2$ and $\delta$ are drawn from their full conditional probabilities; and the Metropolis-Hastings algorithm with a truncated normal proposal is applied to update $\lambda$ \citep{BayesianNetworkMarker2020}. 
As MALA requires the posterior to be differentiable, we approximate the non-differentiable components of the posterior by
\begin{align}
	&\mathbf{1}_{\{\Vert\valpha_{\vi}\Vert_2^2 > \lambda\}} \approx t(\valpha_{\vi};\lambda) = \frac{1}{2} + \frac{1}{\pi}\arctan{\left(\frac{\Vert\valpha_{\vi}\Vert_2^2 - \lambda}{\epsilon_0} \right)}, \label{smoo_indi} \\
	&\sum_{(\vi,\vi') \in \sE}\Vert\valpha_{\vi}-\valpha_{\vi'}\Vert_2 \approx\sum_{(\vi,\vi') \in \sE}\sqrt{\Vert\valpha_{\vi}-\valpha_{\vi'}\Vert_2^2 + \epsilon_1}. 	\label{smoo_fus}
\end{align}
The approximations become exact if the parameters $\epsilon_0$ in \eqref{smoo_indi} and $\epsilon_1$ in \eqref{smoo_fus} go to $0^+$. 

Though random walk metropolis does not rely on the assumption of smooth posterior, its low efficiency makes it impractical to apply in high-dimensional problems. We compare MALA and the random walk metropolis in Section \ref{supp:compare_RW} of the Appendix through a simulation experiment. The experiment demonstrates the advantage of MALA, and it is worthwhile to smooth the likelihood and prior.
The idea of approximating the non-differentiable thresholding function and $\ell_1$-norm by smooth ones is commonly used in many areas such as spiking neural networks \citep{bohte2000spikeprop} and brain-machine interface technology \citep{onaran2013sparse}. 
Another advantage of using the smooth approximation is to improve the computational efficiency of MCMC \citep[see, e.g.,][]{rischard2018bias}. Furthermore, our approximations \eqref{smoo_indi} and \eqref{smoo_fus} can be interpreted as Student $t$ smoothing with $1$ and $2$ degrees of freedom, respectively. This is similar to the Gaussian smoothing technique of \cite{chatterji2020langevin}. 
Details of these smoothing representations are provided in Section~\ref{supp::t_smoothing} of the Appendix. 

Algorithm~\ref{supp:MALA} in Section~\ref{supp:Posterior} of the Appendix presents the details of the posterior updates.
 With the training sample size $N$, the computational complexity of our algorithm is $O(NpK+pK^2)$, where $p$ is the number of entries of the tensor covariate (i.e., $p=P_1P_2\cdots P_D$ for a $D$-way tensor) and $K$ is the dimension of the spline bases.
After the algorithm execution,  the active regions are determined by the estimated receiver operating characteristic (ROC) curve \citep{hajian2013receiver} according to the posterior samples of BFEN, which is also provided in Section~\ref{supp:Posterior} of the Appendix. The posterior point estimator $\widehat f_{\vi}$ of the additive component function in the active regions is computed by $\vphi\trans\widehat\vbeta_{\vi}$ where $\widehat\vbeta_{\vi}$ is the posterior mean of the truncated spline coefficients \eqref{sparse_dec}. 
Overall, our method includes several hyperparameters $\{r, \rho_{\valpha}, p_0, p_1, \sigma^2_{\mu}, p, a, b\}$ and tuning parameters $\{\delta', \epsilon_0, \epsilon_1\}$ in \eqref{prior_mu_lambda}--\eqref{smoo_fus}. 
For ease of tuning, we suggest to standardize the responses in practice.
After this, we assign $(2P_1\cdots P_D - 1)K/2$ to $p_1$ as discussed in Section~\ref{Baysian_model} and a small number $10^{-6}$ to $\epsilon_1$. 
The choice of $\epsilon_0$ is data-driven and addressed in Section~\ref{supp:para_indi} of the Appendix. 
 We find that our model is not sensitive to the specific choice of small value for $\epsilon_1$ through a sensitivity analysis in Section~\ref{supp:para_indi} of the Appendix. 
We also suggest to set $\delta' = 0.0001$ in prior \eqref{naive_p}, and set $\sigma^2_{\mu} = 100$, $p = 1$ and $a = b = 0.5$ in prior \eqref{prior_mu_lambda}.
 The sensitivity analyses of these parameters are presented in Section~\ref{sens_ana} of the Appendix. 
As for $(r,\rho_{\valpha})$ in the prior~\eqref{naive_p} and $p_0$ in the hyperprior \eqref{prior_del_sig}, a validation method is suggested since they are critical in controlling the strength of the prior. 
In our experiments, we split the available data into a training set and a validation set with sizes in the ratio of $5$ to $1$, and the optimal parameters are those minimizing the validation loss $\text{L}(\vy_{\text{valid}}, \widehat \vy_{\text{valid}}) := ({1}/{N_{\text{valid}}}) \Vert\vy_{\text{valid}} - \widehat \vy_{\text{valid}}\Vert_2^2$, where $N_{\text{valid}}$ is the size of the validation set, and $\widehat\vy_{\text{valid}}$ is the vector of predicted values of the observations $\vy_{\text{valid}}$ in the validation set. 
The details of this procedure are discussed in Section~\ref{supp:validation} of the Appendix.
We find that the above strategy of selecting the hyperparameters works reasonably well in all of our numerical experiments. 

\section{Simulation}\label{simulation}
In this section we compare our method, Bayesian additive tensor regression with FEN prior (BFEN), with three alternative methods: i) the sparse nonparametric tensor additive regression (STAR) with the group lasso penalty \citep{hao2021sparse};
ii) the frequentist linear tensor regression (FTR) with the lasso penalty \citep{zhou2013tensor}; iii) the Bayesian linear tensor regression \citep[BTR,][]{BTR2017}. 
	
\subsection{Simulation Settings}\label{simu_setting}

In our simulation study, the covariate $\mX$ is a $2$-way tensor (i.e., matrix) of dimension $P_1\times P_2$, and so the corresponding additive model can be written as $f(\mX) = \mu + \sum_{i,j}f_{ij}(X_{ij})$ where many $f_{ij}$'s are identically zero.

We let $\mu = 0$ and consider three different patterns of true active regions (non-zero additive component functions):
low-rank shapes, a horse shape, and a shape of handwritten Arabic six from MNIST database \citep{lecun1998mnist}. These patterns are depicted in 
Figure~\ref{strong_simu} where the non-black pixels indicate the positions of the active regions.
	
Each pattern includes two nonlinear settings with different levels of signal-to-noise ratio $\mathrm{SNR}=5$ and $\mathrm{SNR}=50$ respectively, and one linear setting with $\mathrm{SNR}= 5$. 
In the nonlinear settings, for each pixel 
$(i,j)$ in the true active regions, we set $f_{ij} (x) = h_{ij}(x)-m_{ij}$ with
\begin{equation} \label{eqn:simuGfun}
  h_{ij}(x) = a_{ij}\sin(c_{ij}x)+a_{ij}\cos(d_{ij}x)+b_{ij}x,
\end{equation}
and $m_{ij}  = \int h_{ij}(x)\intd x$ such that $f_{ij}(x)$ is centered. 

\begin{table}[t]
	\centering
	\caption{
	Specification of  the component function coefficients $a_{ij}$, $b_{ij}$, $c_{ij}$, and $d_{ij}$ in~\eqref{eqn:simuGfun} for $(i,j)$ in the true active regions for each simulation setting. The nine settings are organized into three groups by their patterns (shapes) of the active regions.
	}
	\resizebox{0.97\textwidth}{!}{	\begin{tabular}
 { |c|  C{1.8cm}| C{1.8cm}|  C{1.8cm} |   C{1.8cm} |C{1.8cm}| C{1.8cm}|  C{1.8cm} |C{1.8cm} | C{1.8cm}|}
		\hline
		Setting ID & 1 & 2  & 3 & 4 & 5   & 6 & 7 & 8 & 9\\
		\hline
		Shape & \multicolumn{3}{c|}{Low rank} & \multicolumn{3}{c|}{Horse} & \multicolumn{3}{c|}{Handwritten Arabic six}\\
  		\hline 
    		SNR & $5$ & $50$ & $5$& $5$ & $50$ & $5$& $5$ & $50$ & $5$\\
      \hline 
        \multirow{2}{*}{Setting Meaning}& low SNR & high SNR & \multirow{2}{*}{linear} & low SNR & high SNR & \multirow{2}{*}{linear} & low SNR & high SNR & \multirow{2}{*}{linear} \\
                   & nonlinear &  nonlinear & & nonlinear &  nonlinear & & nonlinear &  nonlinear & \\
		
		\hline
		True $f_{ij}$ & \multicolumn{9}{c|}{$ a_{ij}\sin(c_{ij}x)+a_{ij}\cos(d_{ij}x)+b_{ij}x - m_{ij}$}\\
		\hline
		$m_{ij}$ & \multicolumn{9}{c|}{$\int a_{ij}\sin(c_{ij}x)+a_{ij}\cos(d_{ij}x)+b_{ij}x\intd x$ }\\
		\cline{2-10} 		
		$a_{ij}$ & \multicolumn{2}{c|}{$1$}        & $0$ & \multicolumn{2}{c|}{   $1\overline{u}_{ij}^{(1)}+2$} & $0$  &  \multicolumn{2}{c|}{$2W_{ij} + 1$}   & $0$\\
		$c_{ij}$ &  \multicolumn{2}{c|}{$1.5\pi$}    & $0$ &  \multicolumn{2}{c|}{$v_{ij}^{(2)}$}   & $0$  &  \multicolumn{2}{c|}{ $v_{ij}^{(2)}$}   & $0$\\
		$d_{ij}$ &  \multicolumn{2}{c|}{$1.5\pi$}   & $0$ & \multicolumn{2}{c|}{ $v_{ij}^{(3)}$} & $0$  &  \multicolumn{2}{c|}{$v_{ij}^{(3)}$}   & $0$\\
		$b_{ij}$ &  \multicolumn{2}{c|}{$\frac{2}{\pi}a_{ij}(c_{ij} + d_{ij})$}  & $1$ &  \multicolumn{2}{c|}{$\frac{2}{\pi}a_{ij}(c_{ij} + d_{ij})$}   & $1$  & \multicolumn{2}{c|}{ $\frac{2}{\pi}a_{ij}(c_{ij} + d_{ij})$}   & $1$\\
        \hline
\end{tabular}}
	\label{table1}
\end{table}
We now specify the additive component functions in the active regions through the coefficients $a_{ij}$, $b_{ij}$, $c_{ij}$, and $d_{ij}$ in~\eqref{eqn:simuGfun} for each setting. First, for the linear cases, we let $a_{ij}=0$ and $b_{ij} = 1$ in all three patterns. 
For the nonlinear cases, we set $a_{ij}$ for three patterns in different ways. 
In particular, $a_{ij}$ is set to $1$ for every pixel $(i,j)$. 
For the shape of handwritten Arabic six, we let $\mW$ be the gray-scale matrix of this figure in the MNIST database, and 
 $a_{ij}$ is set as $2W_{ij} + 1$.
For the horse shape, we follow \cite{dong2016learning} which applies the eigenvectors of the graph Laplacian matrix to produce smooth signals on the graph. More specifically, we construct the spatially smooth coefficients $a_{ij}$'s based on the eigenvectors of the graph Laplacian matrix of the graph $\sG$ defined in Section~\ref{Tensor_Additive}. As for $c_{ij}$ and $d_{ij}$, we set them to $1.5\pi$ in the nonlinear cases of the low-rank shapes. For the other two shapes, $c_{ij}$ and $d_{ij}$ are also spatially smooth with value restricted to $[\pi, 1.5\pi]$. Then, $b_{ij}$ is set as $({2}/{\pi}) a_{ij}(c_{ij} + d_{ij})$ for all the nonlinear settings. Finally, we generate the noise terms by adjusting the variance to achieve $\text{SNR} = 50$ for Settings 2, 5, 8, and $\text{SNR} = 5$ for the others. 
Overall, we have nine simulation settings with different shapes of active regions, signal-to-noise ratios, and complexities of the nonlinear functions. These nine settings are summarized in Table~\ref{table1} where the details of constructing $\overline{u}_{ij}^{(1)}$, $v_{ij}^{(2)}$, and $v_{ij}^{(3)}$ by following \cite{dong2016learning} are provided in Section~\ref{supp:setting_simu} of the Appendix. 
	
For each setting, the entries of each covariate $\mX$ are generated from i.i.d.~$\text{unif}(0,1)$, and the response is generated from the additive model with corresponding observational noise level $\sigma_{\epsilon}^2$.
We generated 30 simulated datasets of sample size $600$ independently for each setting.
We apply the proposed BFEN and the alternatives on the datasets.
For BFEN, the hyperparameters are selected as discussed in Section \ref{MALA_Warmstart}. 
For STAR, FTR and BTR, we implement these three methods respectively following \cite{hao2021sparse}, \cite{zhou2013tensor} and \cite{BTR2017}, and the details are provided in Section~\ref{supp:para_exp} of the Appendix. 

To evaluate the estimation accuracy of the component functions for various methods, we calculate the mean squared error (MSE) and relative mean squared error (RMSE) as
$$
    \text{MSE} = \frac{1}{P_1P_2}\sum_{i,j}\Vert f_{ij} - \widehat f_{ij}\Vert_{\bbL_2}^2 \quad \text{and} \quad \text{RMSE} = \frac{1}{|\sV|}\sum_{(i,j)\in\sV} \Vert f_{ij} - \widehat f_{ij}\Vert_{\bbL_2}^2 / \Vert f_{ij}\Vert^2_{\bbL_2},
$$
where $\sV$ is the set of indices of true active functions.
The ability to select the active functions/pixels is assessed by the true positive rate (TPR) and the true negative rate (TNR). 
Note that the posterior samples of the BTR method do not directly indicate the activity of pixels directly. 
To evaluate the region selection performance of BTR, we follow \cite{BTR2017} to identify the active pixels of BTR by checking whether the $95\%$ posterior credible intervals exclude $0$.
For our proposed BFEN method, we used the posterior sample as introduced in Section \ref{MALA_Warmstart}. We further use the testing relative prediction error (RPE) to evaluate the prediction accuracy. 
To do this, we generate another $400$ observations as a testing dataset whose index set is denoted by $\sT$, and calculate
\begin{align}\label{rpe_fomu}
    \text{RPE} = \sum_{n\in\sT} (\widehat{y}_n - y_n)^2 /  \sum_{n\in\sT} y_n^2,
\end{align}
where $\widehat y_n$ the predicted value of the $n$-th observation in the test set through $\widehat\mu$ and $\widehat f_{ij}$'s.
	
\subsection{Results}\label{simulation_result}
 
\begin{figure}[h!]
	    \centering
	    \includegraphics[width=\textwidth]{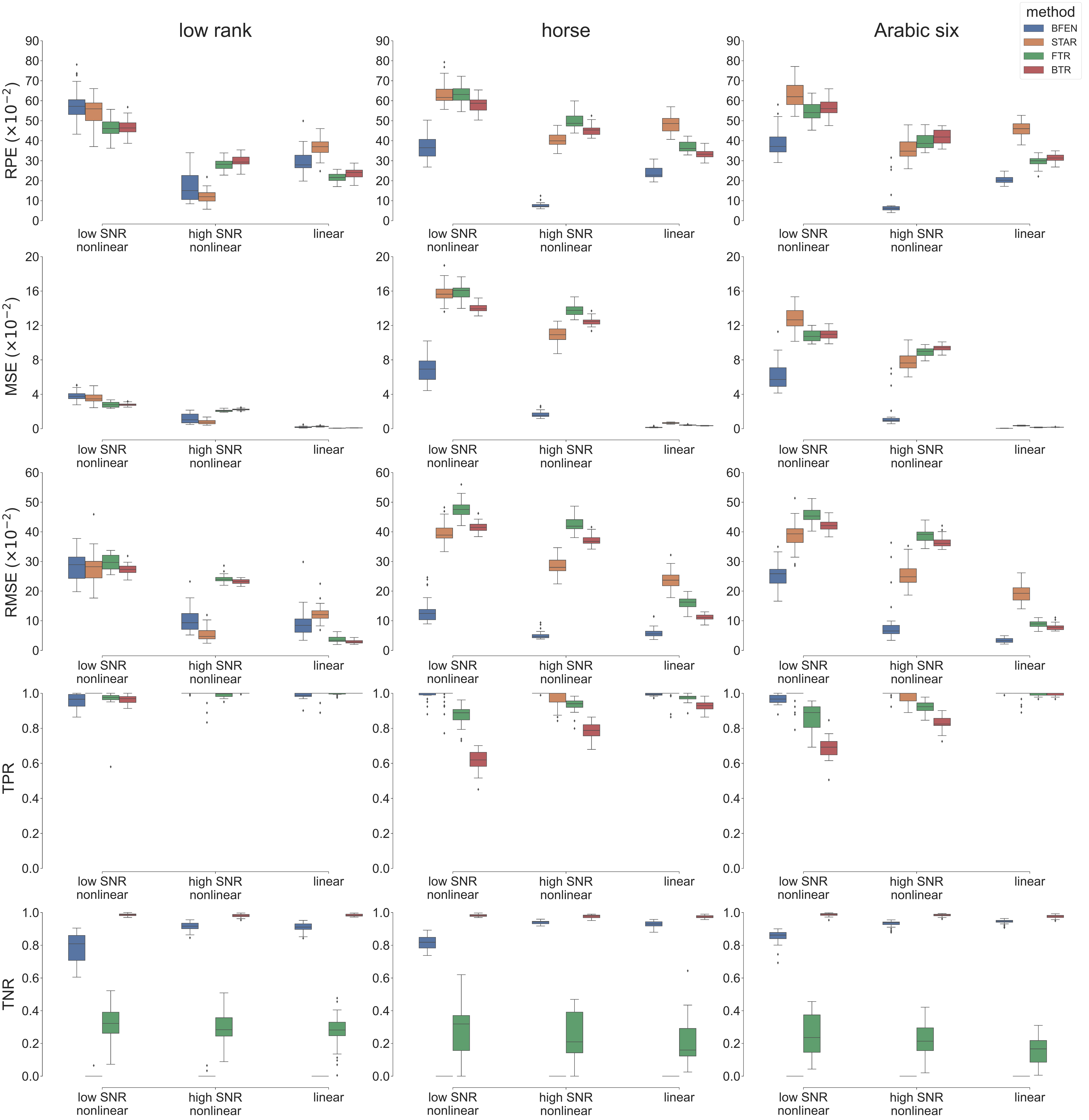}
		\caption{ The boxplots for visualizing the results of simulation experiments. Rows 1--5 depict RPE, MSE, RMSE, TPR, and TNR, respectively. Columns 1--3 respectively correspond to low-rank shapes, a horse shape, and a shape of handwritten Arabic six. In each panel, the left, middle, and right group of boxes correspondingly represent the results under `low SNR, nonlinear', `high SNR, nonlinear' and `linear' setting.
  In each setting, the blue, orange, green, and red boxes correspond to BFEN, STAR, FTR and BTR, respectively.}
		\label{Visual_simu}
	\end{figure}

The results are presented visually as boxplots in Figure \ref{Visual_simu}, which summarizes RPE, MSE, RMSE, TPR, and TNR based on 30 replicates for each setting. We also provide detailed numerical results of the simulation experiments in Table \ref{supp:table_simu} of the Appendix. To compare the computational efficiency between our algorithm and the alternatives, all methods were run on the same platform with a 2.2-GHz Intel E5-2650~v4 CPU 
and the execution time is also recorded in Table~\ref{supp:table_simu}. The convergence time of Algorithm~\ref{supp:MALA} for our proposed BFEN method is less than $1.5$ minutes on average for a single specification of tuning parameters.

It can be seen that the average RPE, MSE, and RMSE of BFEN are smaller than those of STAR, FTR, and BTR in all settings of irregular sparsity shapes, i.e., a horse and a handwritten Arabic six (Settings 4--9). 
In Settings 1--3, the true active region is of low rank which is indeed in favor of the other alternative methods.
It is expected that STAR works well in Setting 2 and the linear alternatives have better performance in Setting 3 since the corresponding settings favor these models. 
Besides, among the three alternative methods, STAR enjoys an advantage over FTR and BTR only when nonlinear signals are strong enough (Settings 2, 5, and 8). Overall, BFEN is more flexible and has advantages in a wider range of scenarios.  

As for the recovery of active regions, the proposed BFEN method has a balanced performance in both TPR and TNR for all settings.
We find that STAR and FTR tend to over-select active pixels, i.e., TNR is low. 
On the other hand, TPR of BTR deteriorates considerably when its low rank and linear assumption are violated in Settings 4, 5, 7, and 8. 

\begin{figure}[h!] 
		\centering
		\includegraphics[width=\textwidth]{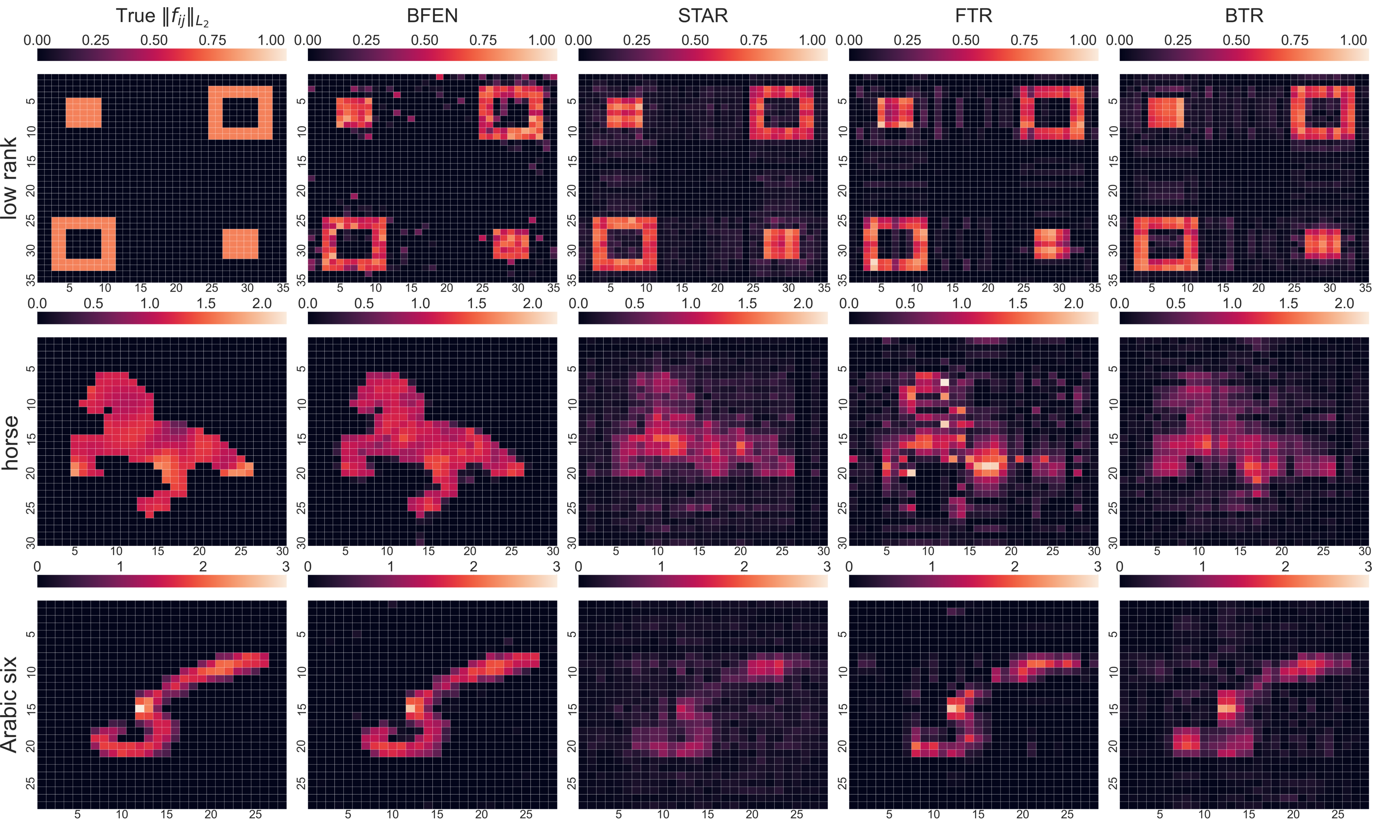} 
		\caption{The heatmaps of various methods under the nonlinear with high SNR settings (Settings 2, 5, and 8). 
		Rows 1--3 correspond to the patterns of low-rank shapes (Setting 2), a horse shape (Setting 5), and a shape of handwritten Arabic six (Setting 8), respectively. 
 		The first column presents the truth. Columns 2--5 correspond to the estimated results by BFEN, STAR, FTR, and BTR, respectively. }\label{strong_simu}
	\end{figure}

For further demonstration, we calculate the $\bbL_2$-norm of each estimated additive component function, $\|\widehat{f}_{ij}\|_{\bbL_2}$, for all methods. These results can be visualized by heatmaps for each simulated dataset. 
For the  nonlinear with high SNR settings, the heatmap corresponding to the median RPE among 30 simulated datasets for each method was depicted in Figure~\ref{strong_simu}, and the heatmap for the truth was also depicted at the leftmost of Figure~\ref{strong_simu}. 
 For the nonlinear with low SNR and linear settings, the heatmaps were respectively provided in Figures~\ref{supp:weak_simu} and \ref{supp:linear_simu} of the Appendix. 
 It is evident that the proposed BFEN recovers the shape of the true active region and the corresponding spatial distribution of the signal strength with a reasonably good accuracy.

In contrast, STAR, FTR, and BTR only work well in the low-rank setting (Setting 2, the first row in Figure~\ref{strong_simu}) but are substantially worse for the other two patterns. 
STAR, FTR, and BTR are based on the tensor rank-$R$ CP decomposition, which is the sum of $R$ rank-$1$ tensors. Therefore, the sparsity patterns recovered by these methods tend to be a combination of several rectangular blocks. 
BFEN, however, encourages the similarity of neighbouring signals rather than enforcing certain shapes of spatially connected regions and, thus, can adaptively identify the active regions with complex shapes. 
     
\section{Facial Feature Analysis}\label{real_application}
	
We apply our method to the Labeled Faces in the Wild dataset \citep{huang2008labeled}. This dataset consists of facial images collected from $5,721$ people and attributes that quantify various facial features 
for each facial image \citep{kumar2009attribute}. In this experiment, we select one facial image per person and choose
the facial expressions related to the mouth as responses, which are {\it smiling}, {\it frowning}, {\it mouth closed}, {\it mouth wide open}, and {\it teeth not visible}.

   	\begin{figure}[h!]
		\centering
		\includegraphics[width=0.9\textwidth]{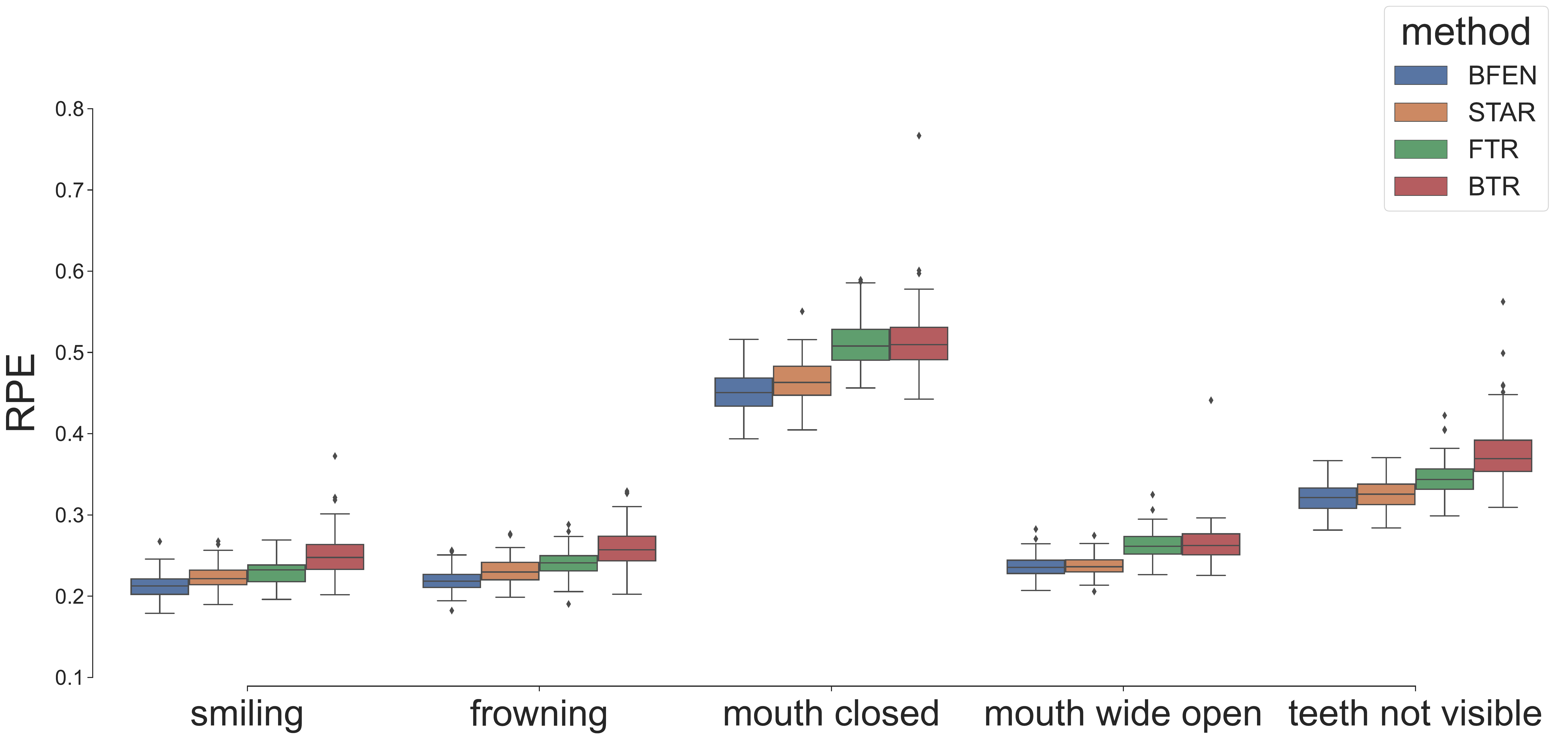}
		\caption{ The boxplots of the relative predictive errors for the facial data analysis under 100 replicates. From left to right, the 5 groups of the boxes respectively represent the results for attributes  {\it smiling}, {\it frowning}, {\it mouth closed}, {\it mouth wide open}, and {\it teeth not visible}, respectively. In each group, the blue, orange, green, and red boxes respectively correspond to BFEN, STAR, FTR, and BTR.}
		\label{Visual_real}
	\end{figure}

\begin{figure}[h!]
		\centering
		\subfigure[]{
			\includegraphics[width=0.25\textwidth]{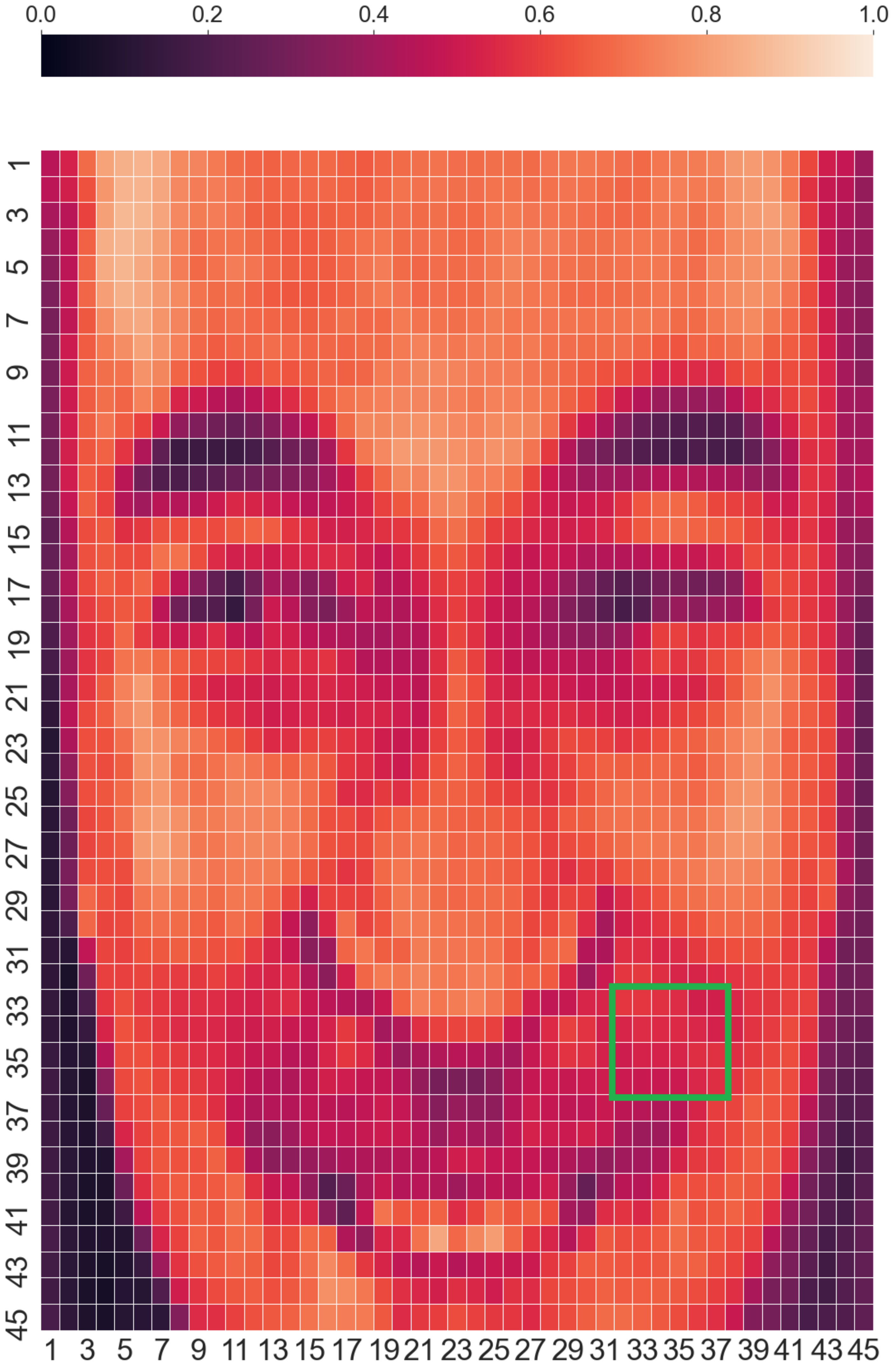}\label{real_frowning_face}
		}\subfigure[]{
			\includegraphics[width = 0.73\textwidth]{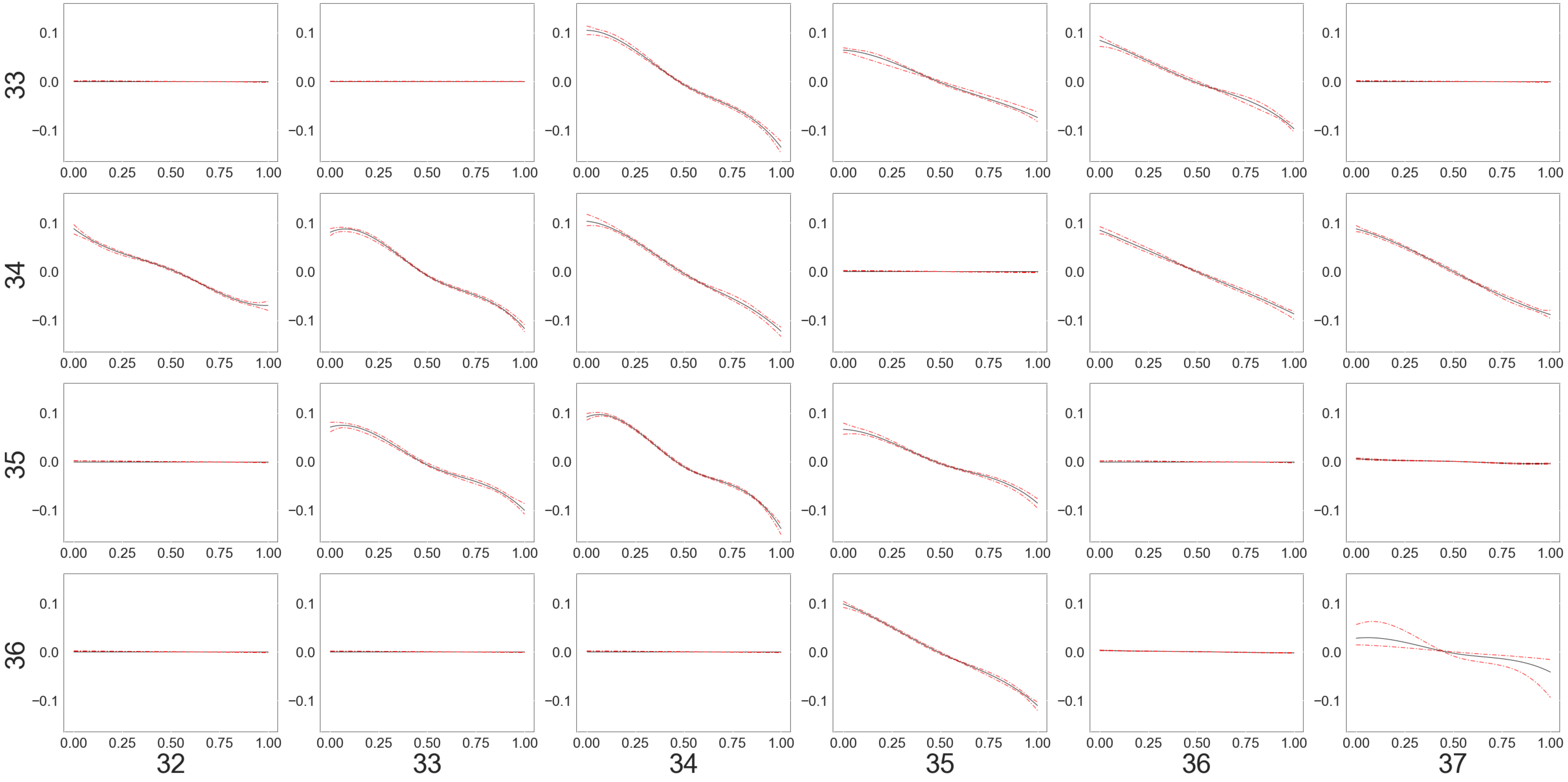} \label{real_frowning_fun}
		}
		\subfigure[]{
			\includegraphics[width=\textwidth]{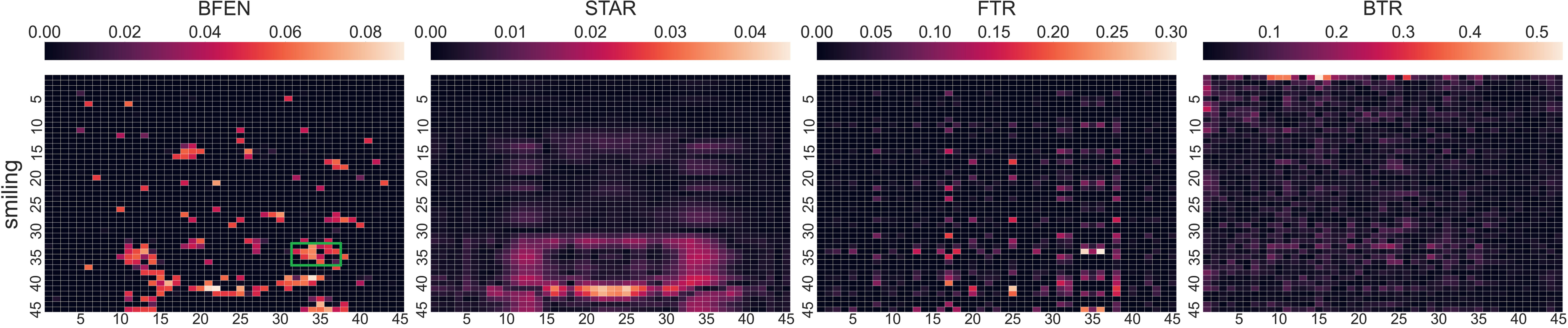} \label{real_frowning_heat}
		}
		\caption{Real applications on the facial data for the attribute {\it smiling}. (a) An example of facial covariate tensor $\mX$. Its corresponding value of smiling attribute is $1.51$, which means the person is smiling. (b) Each of the $4\times 6$ panels depicts the estimated $\widehat f_{ij}$ from BFEN corresponding to the enclosed rectangle area in (a), and between the dashed lines are the $95\%$ posterior credible intervals. (c) The heatmaps in columns 1--4 correspond to $\Vert\widehat f_{ij}\Vert_{\bbL_2}$ estimated by BFEN, STAR, FTR, and BTR, respectively.
		}
		\label{real_frowning}
	\end{figure}

We follow \cite{hassner2015effective} to register these images.
In particular, 
all images are frontalized
to make faces in constrained and forward-facing poses; thus, the same regions of different images represent the same part of a human face. 
The original gray-scale image is of size $90\times 90$ with entry values in $[0,255]$. We further down-sample each image to a $45\times 45$ matrix by replacing every four pixels in a square with one pixel of average gray-scale value, and rescale the entry values to $[0,1]$. Figure~\ref{real_frowning_face} shows an example of the resulting image. 

We compare the proposed method with STAR, FTR, and BTR as in Section~\ref{simulation}.
We randomly sample an index set $\sS$ of size $2000$ from the full subject set $\{1,\cdots,5721\}$  for feasible computation.
The set $\sS$ is then divided into three disjoint subsets $\sS = \sS_1\cup\sS_2\cup\sT$ of sizes $1000$, $200$ and $800$ respectively. 
Sets $\sS_1$ and $\sS_2$ are used for training and tuning, and Set $\sT$ is used to evaluate the performance of prediction through RPE in \eqref{rpe_fomu}. We repeat this procedure 100 times. 

The results are presented visually as boxplots in Figure \ref{Visual_real}, which summarizes the RPE of various methods for each attribute. We also provide the numerical results and runtime for the facial feature analysis in Table~\ref{supp:table_real_RPE} of the Appendix. 
In particular, Algorithm~\ref{supp:MALA} of our proposed BFEN method needs less than $2$ minutes on average to converge for one grid of tuning parameters with a 2.2-GHz Intel E5-2650 v4 CPU. 
It shows that BFEN outperforms the three competitors in all cases, except for the response {\it mouth wide open} where BFEN and STAR have similarly good performances.
The heatmaps in Figure~\ref{real_frowning_heat} display the magnitude $\|\widehat{f}_{ij}\|_{L_2}$ of each pixel for the attribute {\it smiling} using various methods. 
It shows that the result of BFEN has better interpretability: {\it smiling} can be characterized by the pixel values around the eyes, mouth and some facial muscles.
Figure~\ref{real_frowning_fun} depicts the estimated nonlinear functions $f_{ij}$'s and the 95\% posterior credible intervals by BFEN corresponding to the region indicated by the rectangle in Figure~\ref{real_frowning_face}. 
Some functions exhibit clear non-linearity. 
In contrast, the signals selected by FTR and BTR do not have an obvious interpretation. 
With the help of nonlinearity and the group regularization across different blocks, STAR has better interpretability than that of FTR and BTR, but is still inferior to BFEN. 
Overall, the low-rank modeling may not be flexible enough to characterize a complex shape like smiling, and this result is consistent with our findings in the simulation study.
The heatmaps for other attributes are depicted in Figure~\ref{supp:real1} of the Appendix.
	
\section{Discussion}\label{sec:dis}	

In this paper, we have proposed a nonlinear Bayesian tensor additive regression model, which incorporates the spatial information of the tensor covariates. A functional version of the fused elastic net, FEN, has been introduced as a prior distribution on the additive component functions to accommodate the sparse, spatially smooth functional structure with discontinuous jumps.
Through numerical experiments on the simulated and the facial feature datasets, we have demonstrated the superior performance of the proposed method compared to the existing linear and nonlinear tensor regression models  for characterizing irregular shapes of sparse active regions, even if the signal-to-noise ratio is relatively low. The performance of alternative methods, however, rely on low-rank assumption, which is often violated in real applications of image and neuroscience data. 

The proposed BFEN has some limitations, which may lead to extension of this work. Similar to many other methods with multiple hyperparameters, the main computational burden of our method is due to the validation method for selecting hyperparameters $p_0$, $r$, and $\rho_{\valpha}$. Further investigation is needed to relieve this bottleneck by, for example, imposing appropriate hyperpriors on these hyperparamters to automatically adjust them. 
In addition, extending the current model to the case of multi-dimensional response variables, like matrix-on-tensor and tensor-on-tensor regressions, is also of interest.

\spacingset{1.05}
\newpage
\if1\blind
{
  \bigskip
  \bigskip
  \bigskip
  \begin{center}
    {\LARGE\bf Appendix}
\end{center}
  \medskip
} \fi		

	\setcounter{equation}{0}
	\setcounter{theorem}{0}
	\setcounter{lemma}{0}
	\setcounter{proposition}{0}
	\setcounter{section}{0}
	\setcounter{table}{0}
	\setcounter{figure}{0}
	\renewcommand{\theequation}{A.\arabic{equation}}
	\renewcommand{\thetheorem}{A.\arabic{theorem}}
	\renewcommand{\theproposition}{A.\arabic{proposition}}
	\renewcommand{\thelemma}{A.\arabic{lemma}}
	\renewcommand{\thecondition}{A.\arabic{condition}}
	\renewcommand{\thetable}{A.\arabic{table}}
	\renewcommand{\thetable}{A.\arabic{table}}
	\renewcommand{\thesection}{A.\arabic{section}}
	\renewcommand{\thefigure}{A.\arabic{figure}}
	\renewcommand{\thealgocf}{A.\arabic{algocf}}
	
	
	\allowdisplaybreaks
	\spacingset{1.2}

	\section{Construction of Spline Basis} \label{supp:app:spline}
	In this section, we show the details of the basis construction in Section~\ref{Spl_Repre}. The $K$ dimensional centered and orthonormal  basis $\vphi$ is from an $K+1$ dimensional B-spline basis $\vpsi$. The construction is divided into three steps:
	
	First, denote $\mW := \int \vpsi(x)\vpsi(x)\trans\intd x$. Suppose its eigendecomposition is $\mW=\mV\mGamma_1\mV\trans$, where $\mGamma_1$ is diagonal containing the eigenvalues and $\mV$ is orthonormal containing the eigenvectors in its columns. Set $\tilde \vpsi(x) := \mGamma_1^{-{1}/{2}}\mV\trans \vpsi(x)$ to get an orthonormal basis
	satisfying
	$$\int \tilde \vpsi(x)\tilde\vpsi(x)\trans  \intd x=\mI_{K+1}.$$ 
	
	Next, denote $\vd :=\int \tilde \vpsi(x) \intd x\in \mathbb{R}^{K+1}$, set  $\mT\in\mathbb{R}^{K+1,K}$ as the full column rank matrix with columns orthornormal to $\vd$, i.e., $\mT\trans \vd=\vzero$. Set $\mathbf{\tilde\vphi}(\cdot):=\mT\trans \tilde\vpsi(x)$, and  we get an set of  orthonormal and centered basis functions satisfying
	$$\int \tilde\vphi(\cdot)\tilde\vphi(\cdot)\trans  \intd x=\mT\trans  \mT=\mI_{K}\quad \mbox{and} \quad \int \mathbf{\tilde\vphi}(x) \intd x = \vzero.$$
	
	Finally, denote $\mOmega_0 := \int \tilde\vphi''(x)\tilde\vphi''(x)\trans\intd x$. Suppose it has the eigendecomposition
	$\mOmega_0 = \mU\mGamma_2\mU\trans$, where $\mGamma_2$ is a diagonal matrix of eigenvalues arranged in an increasing order.  
	Set $\vphi(x) := \mU\trans\tilde\vphi(x)$. We can see that $\vphi$ is a centered and orthornormal basis with a diagonal $$\mOmega := \int \vphi''(x)\vphi''(x)\trans\intd x = \mU\trans\mOmega_0\mU = \mU\trans \mU\mGamma_2\mU\trans\mU = \mGamma_2.$$ 
	Hence the properties (i), (ii), and (iii) in Section~\ref{Spl_Repre} of the main paper are all satisfied by the constructed spline basis $\vphi$.
	
	As $\mOmega_0$ is computed  from the second order derivative of $\tilde\vphi$, its smallest eigenvalue is  zero, i.e., $(\mGamma_2)_{11} = 0$. This eigenvalue corresponds to an eigenvector $\vu_1$ such that $\vu_1\trans \tilde\vphi$ is a linear function. 
	This implies the first component of $\vphi$ is a linear function. Because the properties (i) and (ii) in Section~\ref{Spl_Repre} are both satisfied, it is easy to see that $\phi_k\perp\ell[0, 1], k > 1$. Meanwhile, it can be verified that $\phi_1\valpha_{\vi 1}$ is the projection of the function $g_{\vi} = \vphi\trans\valpha_{\vi}$ onto $\ell[0,1]$, i.e., $\mathcal{P} g_{\vi} = \vphi_{1}\valpha_{\vi 1}$, where $\valpha_{\vi 1}$ is the first element of $\valpha_{\vi}$. Hence, with the basis $\vphi$, the roughness norm  $\sR(g_{\vi}) =\Vert g''_{\vi}\Vert_{\bbL_2}^2 +\delta'\| \mathcal{P} g_{\vi}\|_{\bbL_2}^2$ in~(\ref{eqn:ffenp}) of the main paper can be rewritten as $\sum_{\vi \in\sI} \valpha_{\vi}\trans\mR \valpha_{\vi}$ where $\mR = \mOmega + \delta'\Vert\phi_1\Vert_{\bbL_2}^2\mathbf{e}_1\mathbf{e}_1\trans$ with $\mathbf{e}_1 = (1, 0, \dots, 0)\trans$.

Using the above procedure, constructing the orthonormal basis only requires to know the degree and dimension of the B-spline. For the degree, it can be fixed as $4$ (cubic spline) to alleviate the computational burden, and this choice is commonly used in nonparametric literature \cite{huang2010variable}. As for $K$, there are many simple recommendations based on the sample size \citep[e.g.,][]{ruppert2003semiparametric}. We follow \cite{fan2011nonparametric} to fix $K = \lceil n^{1/5} \rfloor$, where $\lceil \cdot \rfloor$ denotes rounding to the nearest integer, $n$ is the sample size, and the interior knots are equally-spaced quantiles of all covariate samples.
All the results of numerical experiments exhibited in our paper are obtained through this empirical rule, and we find that this rule works reasonably well.

\section{The Order of Normalizing Term of the Prior}\label{proper_hyp}
	
In this section, we provide the proof of Proposition \ref{cons_thm} by calculating the orders of the normalizing term $C_{\delta, \sigma^2}$ of the prior (\ref{naive_p}) with respect to $\delta$ and $\sigma^2$ respectively. 

The normalizing term $C_{\delta, \sigma^2}$ equals to
\begin{equation}\label{supp::int_prior}
	\begin{aligned}
		&\int_{\mathbb{R}^{P_1\cdots P_D K}}\exp\bigg\{ -\frac{\delta\sum_{\vi\in\sI}\valpha_{\vi}\trans \mR\valpha_{\vi}}{\sigma^2} \\
		& \qquad \qquad \qquad \qquad \qquad - \frac{r\sum_{(\vi,\vi') \in \sE}\Vert\valpha_{\vi}-\valpha_{\vi'}\Vert_2^2 + (1-r)\sum_{(\vi,\vi') \in \sE}\Vert\valpha_{\vi}-\valpha_{\vi'}\Vert_2}{2 \sigma^2\rho_{\valpha}}\bigg\}\intd\valpha
	\end{aligned}
	\end{equation}
	With a formula 
	$$\exp\left(-\frac{\lambda\vert a\vert}{\sigma}\right)=
	\int_0^{\infty}\frac{\lambda}{\sqrt{2\pi\omega^2}}\exp\left(-\frac{a^2}{2\sigma^2\omega^2}-\frac{\lambda^2\omega^2}{2}\right)\intd\omega^2$$ for $(\lambda > 0)$ \citep{andrews1974scale},
	we have
	\begin{equation}\label{supp::L1_to_L2}
	\begin{aligned}
		& \exp\bigg\{-\frac{(1-r)\sum_{(\vi,\vi') \in \sE}\Vert\valpha_{\vi}-\valpha_{\vi'}\Vert_2}{2 \sigma^2\rho_{\valpha}}\bigg\} \\
		&\qquad \qquad = \prod_{(\vi,\vi') \in \sE}\int_0^{\infty}\frac{1-r}{\sqrt{2\pi\omega_{\vi\vi'}^2}}\cdot\exp\bigg\{-\frac{\Vert\valpha_{\vi}-\valpha_{\vi'}\Vert_2^2}{2\cdot 4\sigma^4\rho_{\valpha}^2\omega_{\vi\vi'}^2}-\frac{(1-r)^2\omega_{\vi\vi'}^2}{2}\bigg\}\intd\omega_{\vi\vi'}^2 \, .
	\end{aligned}
	\end{equation}
	Here we introduce some notations to facilitate the proof. Let $\valpha_{\cdot k}$ denote the tensor of dimension $P_1\times\cdots\times P_D$ whose $\vi$-th element, $(\valpha_{\cdot k})_{\vi}$, is $\valpha_{\vi k}$ (the $k$-th element of coefficient vector $\valpha_{\vi}$), $\vi=(i_1, \dots, i_D) \in \sI$ and $k=1,\dots, K$. 
	For a generic $D$-way tensor, we define the operator $\text{vec}(\cdot)$ as its vectorization according to the lexicographical order from its $1$-st to $D$-th mode.
	In other words, $\text{vec}(\sI) := (1, 2, \cdots, \prod_{d = 1}^D P_d)$ as the vectorized form of the index set $\sI$ such that $\vi=(i_1, \dots, i_D)$ is now placed at the $t$-th element of $\text{vec}(\sI)$, where $t = i_1 + \sum_{d=2}^D(i_d - 1)\prod_{d'=1}^{d - 1} P_{d'}$, for any $\vi \in \sI$. Similarly, $\mathrm{vec}(\valpha_{\cdot k})$ is the vectorized $\valpha_{\cdot k}$ such that the $t$-th element of $\mathrm{vec}(\valpha_{\cdot k})$ is $(\valpha_{\cdot k})_{\vi}$ whenever $t = i_1 + \sum_{d=2}^D(i_d - 1)\prod_{d'=1}^{d - 1} P_{d'}$. With the vectoried $\valpha_{\cdot k}$, some terms in the integrand of \eqref{supp::int_prior} can be rewritten in terms of quadratic forms. In particular, define the quadratic matrices $\mLambda_1^{(k)}$ for $k=1,\dots, K$, $\mLambda_2$, and $\mLambda_3^{(\vomega)}$ as follows.
	$\mLambda_1^{(k)}$ is $G_{kk}\cdot\mI$, where $G_{kk} \in \mathbb{R}$ is the $k$-th diagonal element of the matrix $\mathbf{R}$ in \eqref{supp::int_prior}. For any edge $(\vi,\vi')\in\sE$ of the graph, suppose $\vi$ (or $\vi'$) corresponds to the $t$-th (or $t'$-th resp.) element of $\text{vec}(\sI)$, the $(t, t')$-th and $(t, t)$-th elements of $\mLambda_2$ and $\mLambda_3^{(\vomega)}$ are $(\mLambda_2)_{tt'} = -{1}/{(2\rho_{\valpha})}$, $(\mLambda_2)_{tt} = -\sum_{s\neq t}(\mLambda_2)_{ts}$, 
	$(\mLambda_3^{(\vomega)})_{tt'} = -{1}/{(2\cdot4\rho_{\valpha}^2\omega_{\vi\vi'}^2)}$, and $(\mLambda_3^{(\vomega)})_{tt} = -\sum_{s\neq t}(\mLambda_3^{(\vomega)})_{ts}$; The other elements of $\mLambda_2$ and  $\mLambda_3^{(\vomega)}$ are $0$'s. The three matrices satisfy:
	\begin{align}
	    \text{vec}(\valpha_{\cdot k})\trans\mLambda_1^{(k)}\text{vec}(\valpha_{\cdot k}) &= \delta\sum_{\vi\in\sI}G_{kk}\alpha_{\vi k}^2,\label{Lambda1}\\
		\text{vec}(\valpha_{\cdot k})\trans\mLambda_2\text{vec}(\valpha_{\cdot k}) &=\sum_{(\vi,\vi') \in \sE}\frac{(\alpha_{\vi k}-\alpha_{\vi'k})^2}{2\rho_{\valpha}},\label{Lambda2}\\
		\text{vec}(\valpha_{\cdot k})\trans\mLambda_3^{(\vomega)}\text{vec}(\valpha_{\cdot k}) &=\sum_{(\vi,\vi') \in \sE}\frac{(\alpha_{\vi k}-\alpha_{\vi'k})^2}{2\cdot4\rho_{\valpha}^2\omega_{\vi\vi'}^2}.\label{Lambda3}
	\end{align}
	Denote 
	$$g(\vomega) = \prod_{(\vi,\vi') \in \sE}\frac{1-r}{\sqrt{2\pi\omega_{\vi\vi'}^2}}\exp\bigg\{-\frac{(1-r)^2\omega_{\vi\vi'}^2}{2}\bigg\}$$ 
	and 
	$$\mLambda(k, \vomega, \delta, \sigma^2) = \mLambda_1^{(k)} + \frac{r\mLambda_2}{\delta} + \frac{\mLambda_3^{(\vomega)}}{\delta\sigma^2}.$$ 
	After using the above simplified notations, we substitute \eqref{supp::L1_to_L2} back into \eqref{supp::int_prior}, and apply the Fubini's Theorem to have
	\begin{equation*}
		C_{\delta,\sigma^2} = \int_{R_+^{\vert\sE\vert}}g(\vomega)\left[\prod_{k=1}^K\int_{\mathbb{R}^{P_1\cdots P_D}}\exp\left\{ -\text{vec}(\valpha_{\cdot k})\trans\frac{\delta\mLambda(k, \vomega, \delta, \sigma^2)}{\sigma^2}\text{vec}(\valpha_{\cdot k})\right\}\intd\valpha_{\cdot k}\right]\intd\vomega^2.
	\end{equation*}
	Integrating out $\valpha_{\cdot k}$'s, the normalizing term $C_{\delta,\sigma^2}$ becomes
	\begin{equation}\label{nor_con}
		c\int_{R_+^{\vert\sE\vert}}g(\vomega)\bigg(\frac{\sigma^2}{\delta}\bigg)^{P_1\cdots P_DK/2}\left[\prod_{k = 1}^K\sqrt{\det \{ \mLambda(k, \vomega, \delta, \sigma^2) \}}\right]^{-1}\intd\vomega^2.
	\end{equation}
	Hence, the key to compute the degrees of $\delta$ and $\sigma^2$ is to find out the degrees within $\det \{ \mLambda(k, \vomega, \delta, \sigma^2) \}$. Since $\sG = (\sI, \sE)$ is a connected graph, \eqref{Lambda2} and \eqref{Lambda3} equal to $0$ if and only if $\valpha_{\vi k} = \valpha_{\vi' k}, \forall \vi, \vi'\in\sI,$ i.e. $\text{vec}(\valpha_{\cdot k})\propto \mathbf{1}$, where $\mathbf{1}$ is a $P_1\cdots P_D$ dimensional vector with each entry being $1$. Hence $\mLambda_2$ and $\mLambda_3^{(\vomega)}$ are positive semidefinite matrix with rank $P_1\cdots P_D-1$.
	Next we discuss the order of $C_{\delta, \sigma^2}$ with respect to $\delta, \sigma^2$ when they go to both $0$ or $\infty.$ On one hand,
	\begin{enumerate}
		\item[(i)] $\delta \rightarrow \infty.$  With the positive (semi-)definiteness of the matrices, we have
		\begin{equation*}0<\det\big(\mLambda_1^{(k)}\big)\leq\det \{ \mLambda(k, \vomega, \delta, \sigma^2) \} \leq\det\bigg(\mLambda_1^{(k)} + r\mLambda_2 + \frac{\mLambda_3^{(\vomega)}}{\sigma^2}\bigg), \quad \forall \delta > 1.\end{equation*}	
		We can see that $\det \{ \mLambda(k, \vomega, \delta, \sigma^2) \}$ is bounded by two positive constants that is independent to $\delta$, so $\det \{ \mLambda(k, \vomega, \delta, \sigma^2) \}$ is of order $O(1)$ for $\delta\rightarrow \infty.$ Together with \eqref{nor_con}, we know the normalizing term $C_{\delta,\sigma^2}$ is of order $(1/\delta)^{P_1\cdots P_DK/2}$. 
		\item[(ii)] $\sigma^2 \rightarrow \infty.$ Similarly, we have 
		\begin{equation*}
		    0<\det\bigg(\mLambda_1^{(k)} + \frac{r\mLambda_2}{\delta} \bigg)\leq \det \{ \mLambda(k, \vomega, \delta, \sigma^2) \} \leq \det\bigg(\mLambda_1^{(k)} + \frac{r\mLambda_2}{\delta} + \frac{\mLambda_3^{(\vomega)}}{\delta}\bigg), \forall \sigma^2 > 1,
		\end{equation*}	
		so $\det \{ \mLambda(k, \vomega, \delta, \sigma^2) \}$ is of order $O(1)$ for $\sigma^2\rightarrow \infty.$ Combining with \eqref{nor_con}, the normalizing term $C_{\delta,\sigma^2}$ is of order $(\sigma^2)^{P_1\cdots P_DK/2}$. 
	\end{enumerate}
	On the other hand, to compute the degrees of $\delta$ and $\sigma$ when they go to $0$, we apply \cite{grinberg2020notes}'s formula for $n$ dimensional square matrices $\mA,\mB$:
	\begin{equation}\label{supp::A_B}
		\det(\mA + x\mB) = \det(\mA) + \det(\mA)p_{1}(\mA^{-1}\mB)x + \cdots + \det(\mA)p_{n-1}(\mA^{-1}\mB)x^{n-1} + \det(\mB)x^n,
	\end{equation}
	where $\mA$ is an invertible square matrix, and $p_1(\cdot), \cdots, p_{n-1}(\cdot)$ are the sums of all principal minors of order $2,\cdots, n-1$, respectively. 
	\begin{enumerate}
		\item[(iii)] $\delta \rightarrow 0$. After respectively substituting the three variates $\mLambda_1^{(k)}, $ $r\mLambda_2 + {\mLambda_3^{(\vomega)}}/{\sigma^2}$, and $1/\delta$ for $\mA$, $\mB$, and $x$ in formula \eqref{supp::A_B}, it shows that $n$ in \eqref{supp::A_B} turns to be $P_1\cdots P_D$, and $\mA$ is a positive definite matrix. It is easy to see that the right side of the linear combination $r \times \eqref{Lambda2} + \eqref{Lambda3}/\sigma^2$ equal to $0$ if and only if $\valpha_{\vi k} = \valpha_{\vi' k}$, $\forall \vi, \vi'\in \sI$, so $\mB$ is a positive semidefinite matrix with rank $P_1\cdots P_D-1$, and meanwhile $\mA^{-1}\mB$ is also positive semidefinite matrix with rank $P_1\cdots P_D-1$. Hence we have		
		\begin{equation*}
		    \det \{ \mLambda(k, \vomega, \delta, \sigma^2) \} = \det(\mA) + \det(\mA)p_1(\mA^{-1}\mB)/\delta + \cdots + \det(\mA)p_{n-1}(\mA^{-1}\mB)/\delta^{n-1},	
		\end{equation*}		
		where $\det(\mB) = 0$, $\det(\mA) >0 $, and $p_{j}\big(\mA^{-1}\mB\big)> 0$, $j = 1,\cdots,n - 1$. So we know that $\det \{ \mLambda(k, \vomega, \delta, \sigma^2) \}$ is of order $1/\delta^{P_1\cdots P_D-1}$ for $\delta\rightarrow 0.$ Combining with \eqref{nor_con}, the normalizing term $C_{\delta,\sigma^2}$ is of order $(1/\delta)^{K/2}$.
		\item[(iv)] $\sigma^2 \rightarrow 0$. Substitute $\mLambda_1^{(k)} + {r\mLambda_2}/{\delta}$, $ {\mLambda_3^{(\vomega)}}/{\delta}$, $1/\sigma^2$ for $\mA$, $\mB$, $x$ in the formula \eqref{supp::A_B}, respectively. Similarly, we can prove $\det \{ \mLambda(k, \vomega, \delta, \sigma^2) \}$ is of order $(1/\sigma^2)^{P_1\cdots P_D-1}$. Further, according to \eqref{nor_con}, the normalizing term $C_{\delta,\sigma^2}$ is of order $(\sigma^2)^{(2P_1\cdots P_D-1)K/2}$.
	\end{enumerate}
\section{Model Invariance} \label{supp:app:invarant}
In this section, we prove Proposition~\ref{prop:priorinvariant} in Section~\ref{Baysian_model} of the main paper. 
Denote $p(\valpha;\vphi)$ the probability distribution function of the prior (\ref{naive_p}), where `$\vphi$' is involved because $\mR$ in the prior (\ref{naive_p}) is defined based on the spline basis.
Note $f_{\vi}(X_{\vi}) = \vphi(X_{\vi})\trans\valpha_{\vi}\cdot \mathbf{1}_{\{\Vert\valpha_{\vi}\Vert_2^2 > \lambda\}}$, so the prior $p(\valpha;\vphi)$ of paramater $\valpha$ induces a prior distribution for $f(\mX)$. For any orthogonal matrix $\mQ\in O(K)$, after giving orthogonal transformation to the spline basis: $\vphi_{\mQ} = \mQ\vphi,$ we have a new model $Y = f_{\mQ}(X) + \epsilon$ with component function 
\begin{align}\label{supp::f_Qi_1}
    f_{\mQ,\vi}(X_{\vi}) = \vphi_{\mQ}(X_{\vi})\trans\valpha_{ \vi}\cdot \mathbf{1}_{\{\Vert\valpha_{ \vi}\Vert_2^2 > \lambda\}}.
\end{align} 
With the new spline basis $\vphi_{\mQ}$, the prior imposed on $\valpha$ turns to be $p(\valpha;\vphi_{\mQ})$, which also induces a prior distribution for $f_{\mQ}(\mX)$. Proposition \ref{prop:priorinvariant} is an equivalent to: The prior distribution of $f_{\mQ}(\mX)$ induced by the prior $p(\valpha;\vphi_{\mQ})$ keeps unchange for any orthonormal matrix $\mQ\in O(K)$.
Denote $\valpha_{\mQ}\in \bbR^{P_1\times\cdots\times P_D\times K}$ whose element $(\valpha_{\mQ})_{\vi k} = (\mQ\trans\valpha_{\vi})_k$, $\vi\in\sI, 1\leq k \leq K$. Because 
$$
    f_{\mQ,\vi}(X_{\vi}) = \vphi(X_{\vi})\trans\mQ\trans\valpha_{\vi}\cdot \mathbf{1}_{\{\Vert\valpha_{ \vi}\Vert_2^2 > \lambda\}}= \vphi(X_{\vi})\trans\mQ\trans\valpha_{\vi}\cdot \mathbf{1}_{\{\Vert \mQ\trans\valpha_{ \vi}\Vert_2^2 > \lambda\}},
$$ 
\eqref{supp::f_Qi_1} with the prior $p(\cdot;\vphi_{\mQ})$ for $\valpha$ is equivalent to 
\begin{align}\label{supp::f_Qi_2}
    f_{\mQ,\vi}(X_{\vi}) = \vphi(X_{\vi})\trans(\valpha_{\mQ})_{\vi}\cdot \mathbf{1}_{\{\Vert (\valpha_{\mQ})_{\vi}\Vert_2^2 > \lambda\}}
\end{align} 
with a prior $q(\cdot;\vphi_{\mQ})$ for $\valpha_{\mQ}$, where $q(\cdot;\vphi_{\mQ})$ is obtained through density transformation from $p(\valpha;\vphi_{\mQ})$. All we need is to show that $q(\cdot ;\vphi_{\mQ})$ is invariant to $\mQ\in O(K)$, which, is guaranteed by the following computation
\begin{align*}\begin{split}
    \ln q(\valpha; \vphi_{\mQ}) & \propto -\sum_{\vi} \valpha_{\vi}\trans\mQ\trans(\mQ\mR\mQ\trans)\mQ\valpha_{\vi} 
	-\frac{r\sum_{(\vi,\vi')\in\sE}\big(\valpha_{\vi}\trans\mQ\trans\mQ\valpha_{\vi} - 2\valpha_{\vi'}\trans\mQ\trans\mQ\valpha_{\vi} + \valpha_{\vi'}\trans\mQ\trans\mQ\valpha_{\vi'}\big)}{2\sigma^2\rho_{\valpha}} \\
	&\qquad\qquad-\frac{(1 - r)\sum_{(\vi,\vi')\in\sE}\sqrt{\valpha_{\vi}\trans\mQ\trans\mQ\valpha_{\vi} - 2\valpha_{\vi'}\trans\mQ\trans\mQ\valpha_{\vi} + \valpha_{\vi'}\trans\mQ\trans\mQ\valpha_{\vi'}}}{2\sigma^2\rho_{\valpha}}\\
	& \propto  -\sum_{\vi} \valpha_{\vi}\trans\mR\valpha_{\vi} 
	-\frac{r\sum_{(\vi,\vi')\in\sE}\big(\valpha_{\vi}\trans\valpha_{\vi} - 2\valpha_{\vi'}\trans\valpha_{\vi} + \valpha_{\vi'}\trans\valpha_{\vi'}\big)}{2\sigma^2\rho_{\valpha}} \\
	&\qquad\qquad-\frac{(1 - r)\sum_{(\vi,\vi')\in\sE}\sqrt{\valpha_{\vi}\trans\valpha_{\vi} - 2\valpha_{\vi'}\trans\valpha_{\vi} + \valpha_{\vi'}\trans\valpha_{\vi'}}}{2\sigma^2\rho_{\valpha}}\\
    & \propto p(\valpha;\vphi).
\end{split}\end{align*}

\section{Student $t$ Smoothing}\label{supp::t_smoothing}
In this section, we show that the proposed approximations (\ref{smoo_indi}) and (\ref{smoo_fus}) of nonsmooth functions in the main paper can be represented in terms of smoothing method similar to \cite{chatterji2020langevin}, where a Gaussian smoothing was introduced.
	
We first recall that $\valpha\trans = (\valpha_{\vi}\trans)_{\vi \in \sI} \in \mathbb{R}^{P_1\times\cdots\times P_D\times K}$ are the combined spline coefficients for all additive component functions $f_{\vi}$, $\vi \in \sI$, and $\sE$ is the neighboring relationship set for the location index set $\sI$. Let $p$ be the cardinality of $\sE$, i.e., $p = \vert\sE\vert$. 
Denote $\vg: \mathbb{R}^{P_1\times\cdots\times P_D\times K}\rightarrow \mathbb{R}^p$ such that $\vg(\valpha) = (\Vert\valpha_{\vi} - \valpha_{\vi'}\Vert_2)_{(\vi,\vi')\in\sE}$. 
We can then rewrite the non-differentiable fusion term $\sum_{(\vi,\vi')\in\sE}\Vert\valpha_{\vi} - \valpha_{\vi'}\Vert_2$ (the left hand side of (\ref{smoo_fus}) of the main paper) as $\| g(\valpha) \|_1$. Now, let $\vxi$ be a $p$-dimensional random vector with $\xi_{j}\stackrel{\rm i.i.d.}{\sim} t_2$, $ j \in \sE$, where $t_2$ denotes the Student $t$ distribution with 2 degrees of freedom. 
It can be shown that, for $\epsilon_1 >0$, 
\begin{equation}\label{eqn:t2PerturMain}
    \mathbb{E} \,\| g(\valpha) + \sqrt{(\epsilon_1/2)} \, \vxi  \|_1 = \sum_{(\vi,\vi') \in \sE}\sqrt{\Vert\valpha_{\vi}-\valpha_{\vi'}\Vert_2^2 + \epsilon_1}.
\end{equation}
Thus (\ref{smoo_fus}) of the main paper can be represented by a perturbation \citep{chatterji2020langevin} using Student $t$ distribution with 2 degrees of freedom. 
To show \eqref{eqn:t2PerturMain}, we note that for a random variable $\xi\sim t_2$, direct calculation shows
\begin{align}\label{uni_t2perturb}
    \mathbb{E}\vert\alpha + u\xi\vert = \sqrt{\alpha + 2u^2},
\end{align}
for $\alpha,\, u \in R$.
Thus, for $\xi_{(\vi,\vi')}\stackrel{\rm i.i.d.}{\sim} t_2$, we have
\begin{align*}
    \mathbb{E} \,\| g(\valpha) + \sqrt{(\epsilon_1/2)} \, \vxi  \|_1   
    & = \sum_{(\vi,\vi')\in\sE}\mathbb{E} \, \big\vert \Vert\valpha_{\vi} - \valpha_{\vi'}\Vert_2 + \sqrt{(\epsilon_1/2)}\,\xi_{(\vi,\vi')}\big\vert \\
    &  \stackrel{\eqref{uni_t2perturb}}{=}
    \sum_{(\vi,\vi')\in\sE}\sqrt{\Vert\valpha_{\vi} - \valpha_{\vi'}\Vert_2^2 + \epsilon_1},
\end{align*} 
which completes our Student $t_2$ perturbation representation of (\ref{smoo_fus}) of the main paper.

Similarly, let $U$ be the indicator function  such that $U(u) = \mathbf{1}_{\{u > \lambda\}}$, and define $\xi_{\vi}\stackrel{\rm i.i.d.}{\sim} t_1$, $\vi \in \sI$, where $t_1$ denotes the Student $t$ distribution with 1 degree of freedom (i.e., the Cauchy distribution).
We then have
\begin{align*}
    \mathbb{E} \big\{ U \big(\Vert\valpha_{\vi}\Vert_2^2 + \epsilon_0 \xi_{\vi} \big) \big\} & = \mathbb{E}\big(\mathbf{1}_{\{\Vert\valpha_{\vi}\Vert_2^2 + \epsilon_0 \xi_{\vi} > \lambda\}}\big) \\ 
    & = \mathbb{P}\big(\Vert\valpha_{\vi}\Vert_2^2 + \epsilon_0\xi_{\vi} > \lambda \big) \\ 
    &  = \mathbb{P}\big\{ \xi_{\vi} > (\lambda - \Vert\valpha_{\vi}\Vert_2^2)/\epsilon_0 \big\} \\
    &  =1 - \bigg[ \frac{1}{2} + \frac{1}{\pi}\arctan\{(\lambda - \Vert\valpha_{\vi}\Vert_2^2)/\epsilon_0 \} \bigg] \\ 
    &  =\frac{1}{2} + \frac{1}{\pi}\arctan\{(\Vert\valpha_{\vi}\Vert_2^2 - \lambda)/\epsilon_0 \}. 
\end{align*}
Thus, (\ref{smoo_indi}) of the main paper can be represented by the Student $t_1$ (Cauchy) perturbation.

\section{Model Estimation}\label{supp:Model_E}
In this section we demonstrate how to estimate the tensor regression model (\ref{eqn:model}) of the main paper. The method includes two major components: sampling posterior  and selecting  hyperparameters (a validation method).

\subsection{Posterior Sampling}\label{supp:Posterior}	
\begin{algorithm}[htpb]
	\caption{Posterior updates under fixed $r$ and $\rho_{\valpha}$.}
	\label{supp:MALA}
		\SetKwInput{Input}{Input}
		\SetKwInput{Output}{Output}
		\Input{the parameters from the last iteration}
		
		\Output{the updated parameters for the next iteration}
		
		Draw $\mu^*\sim N(\tilde{\mu},\tau_{\mu}^2)$, where
		$$\tilde{\mu}=\mu+\frac{\tau_{\mu}^2}{2}\big(\sum_{n=1}^N\frac{\partial\ln p(y_n\mid\valpha, \mu, \sigma^2, \lambda)}{\partial\mu}+\frac{\partial\ln p(\mu)}{\partial\mu}\big).$$
		Update $\mu=\mu^*$ with probability
		$$\min\left\{1,\frac{N(\mu|\tilde{\mu^*},\tau_{\mu^*}^2)p(\mu^*)\prod_{n=1}^Np(y_n|\valpha, \mu^*, \sigma^2, \lambda)}{N(\mu^*|\tilde{\mu},\tau_{\mu}^2)p(\mu)\prod_{n=1}^Np(y_n|\valpha, \mu, \sigma^2, \lambda)}\right\}.$$
		
		Draw $\valpha^*\sim N(\tilde{\valpha},\tau_{\valpha}^2I_{P_1P_2})$, where
		$$\tilde{\valpha}=\valpha+\frac{\tau_{\valpha}^2}{2}\big(\sum_{n=1}^N\frac{\partial\ln p(y_n\mid\valpha, \mu, \sigma^2, \lambda)}{\partial\valpha}+\frac{\partial\ln p(\valpha\mid \delta, r, \sigma^2, \rho_{\valpha})}{\partial\valpha}\big).$$
		Update $\valpha=\valpha^*$ with probability
		$$\min\left\{1,\frac{N(\valpha|\tilde{\valpha^*},\tau_{\valpha^*}^2I_{P_1P_2})p(\valpha^*\mid \delta, r, \sigma^2, \rho_{\valpha})\prod_{n=1}^Np(y_n|\valpha^*, \mu, \sigma^2, \lambda)}{N(\valpha^*|\tilde{\valpha},\tau_{\valpha}^2I_{P_1P_2})p(\valpha\mid \delta, r, \sigma^2, \rho_{\valpha})\prod_{n=1}^Np(y_n|\valpha, \mu, \sigma^2, \lambda)}\right\}.$$
		
		Draw $\sigma^2\sim \text{Inv-}\Gamma(a,b),$ where $a = p_1 + \frac{N}{2},$ 
		\begin{align*}\begin{split}b = 1 & + \frac{\sum_{n=1}^N (y_n - \mu - \sum_{\vi\in\sI}\vphi(X_{\vi}^{(n)})\trans\valpha_{\vi}\cdot t_{\lambda}(\valpha_{\vi}))^2}{2} + \delta\sum_{\vi\in\sI}\valpha_{\vi}\trans \mR\valpha_{\vi} \\& + \frac{r\sum_{(\vi,\vi') \in \sE}\Vert\valpha_{\vi}-\valpha_{\vi'}\Vert_2^2 + (1-r)\sum_{(\vi,\vi') \in \sE}\Vert\valpha_{\vi}-\valpha_{\vi'}\Vert_2}{2\rho_{\valpha}}.\end{split}\end{align*}
		
		Draw $\delta\sim \Gamma(a,b),$ where $a = p_0$ and $$b = 1 + \frac{\sum_{\vi\in\sI}\valpha_{\vi}\trans\mR\valpha_{\vi}}{\sigma^2}.$$
		
		Draw $\lambda^{*} \sim N_{+}\left(\lambda, 0, \lambda_{u}, \tau_{\lambda}^{2}\right) ,$ which is a normal distribution $N(\lambda, \tau_{\lambda}^{2})$ truncated by $[0, \lambda_{u}]$. Update $\lambda=\lambda^{*}$ with probability
		$$
		\min \left\{1, \frac{N_{+}\left(\lambda | \lambda^{*}, \lambda_{l}, \lambda_{u}, \tau_{\lambda}^{2}\right)p(\lambda^*) \prod_{n = 1}^N f\left(y_{n} | \valpha, \mu, \sigma^2, \lambda^*\right)}{N_{+}\left(\lambda^{*} | \lambda, \lambda_{\lambda}, \lambda_{u}, \tau_{\lambda}^{2}\right)p(\lambda) \prod_{n = 1}^N f\left(y_{n} | \valpha, \mu, \sigma^2, \lambda\right)}\right\}.$$ 
\end{algorithm}	
	
Algorithm~\ref{supp:MALA} describes the Markov chain Monte Carlo (MCMC) method to obtain posterior samples for the parameters $\{\mu, \valpha, \lambda, \sigma^2, \delta\}$ of the hierarchical model (\ref{eqen:likelihood})--(\ref{prior_del_sig}) of the main paper. Steps 1 and 2 use the MALA \citep{roberts1998optimal} to sample $\mu$ and $\valpha$; Step 3 and 4 draw the parameters $\sigma^2$ and $\delta$ from their full conditional probabilities; and Step 5 applies the Metropolis-Hastings algorithm with a truncated normal proposal to update $\lambda$ \citep{BayesianNetworkMarker2020}. 
The hyperparamaters $(r, \rho_{\valpha})$ in~(\ref{naive_p}) and $p_0$ in~(\ref{prior_del_sig}) of the main paper are fixed during the MCMC update.

Given the posterior samples of $\valpha, \lambda$ from Algorithm~\ref{supp:MALA}, we estimate the model coefficient $\vbeta$ in~(\ref{sparse_dec}) of the main paper in the following way.  Denote  $\left\{\valpha^{(B+l)}, \lambda^{(B+l)}\right\}_{l=1}^{I-B}$ the posterior samples after burn-in, and $D_N$ the training dataset. 
We achieve  sparsity by selecting the active indices $(i,j)$ from the posterior inclusion probability.
The posterior inclusion probability for $\vbeta_{\vi}$ is given by the posterior mean of the indicator function $t(\valpha_{\vi};\lambda)$ in~\eqref{smoo_indi}:
$$
	\widehat{\operatorname{Pr}}\left(\vbeta_{\vi} \neq 0 \mid D_{N}\right)=\frac{1}{I - B} \sum_{l=1}^{I - B} t\big(\valpha_{\vi}^{(B + l)};\lambda^{(B + l)}\big).
$$
The corresponding additive component function $f_{\vi}$, $\vi \in \widehat{\sV}$, is regarded as active if $\widehat{\operatorname{Pr}}\left(\vbeta_{\vi} \neq 0 \mid D_{N}\right)>c_0$ for some cut-off value $c_0$. The estimated active index set is then 
$$
	\widehat{\sV}(c_0)=\left\{\vi: \widehat{\operatorname{Pr}}\left(\vbeta_{\vi} \neq \vzero \mid D_{N}\right)>c_0\right\}.$$
Similar to \cite{hajian2013receiver}, the cut-off value $c_0$ can be decided according to the receiver operating characteristic (ROC) curve. For this purpose, we introduce several notations. Define two tensors $\mJ_{\sV}, \mJ_{\widehat{\sV}(c_0)}\in \bbR^{P_1\times\cdots\times P_D}$ such that
$$
    (\mJ_{\sV})_{\vi}=\begin{cases}
	1, & \text{true } \vbeta_{\vi}\neq \vzero,\\
	0, & \text{otherwise;}
	\end{cases} 
	\qquad \text{ and } \qquad
	(\mJ_{\widehat{\sV}(c_0)})_{\vi}=\begin{cases}
	1, & \vi\in\widehat{\sV}(c_0),\\
	0, & \text{otherwise.}
	\end{cases}$$
 In the above, $\vbeta_{\vi}$ is the true coefficient and thus $\mJ_{\sV}$ can be interpreted as an indicator for the true active index, while $\mJ_{\widehat{\sV}(c_0)}$ can be interpreted as the estimated active index.
We also define a tensor	$\mJ\in \bbR^{P_1\times\cdots\times P_D}$ whose elements are all ones, i.e., $\mJ_{\vi} = 1,\ \forall \vi\in\sI$. 
With these notations, the true negative rate (TNR) and the true positive rate (TPR) for the cut-off value $c_0$ are respectively defined as 
$$\text{TNR}(c_0) = \frac{\langle\mJ - \mJ_{\sV}, \mJ - \mJ_{\widehat{\sV}(c_0)}\rangle}{\langle\mJ, \mJ - \mJ_{\sV}\rangle}$$ and 
$$\text{TPR}(c_0) = \frac{\langle\mJ_{\sV},\mJ_{\widehat{\sV}(c_0)}\rangle}{\langle\mJ,\mJ_{\sV}\rangle}.$$ 
Note that $\mathbf{\mJ}_{\sV}$ is unknown, we use the tensor $\mP\in \bbR^{P_1\times\cdots\times P_D}$ whose element $P_{\vi} = \widehat{\operatorname{Pr}}\left(\vbeta_{\vi} \neq \vzero \mid D_{N}\right)$ to approximate $\mathbf{\mJ}_{\sV}$ in practice. 
According to \cite{hajian2013receiver}, the estimation of TNR and TPR can be obtained as
$$
    \widehat{\text{TNR}}(c_0) = \frac{\langle\mJ - \mP, \mJ - \mJ_{\widehat{\sV}(c_0)}\rangle}{\langle\mJ, \mJ - \mP\rangle},\quad \widehat{\text{TPR}}(c_0) = \frac{\langle\mP,\mJ_{\widehat{\sV}(c_0)}\rangle}{\langle\mJ,\mP\rangle}.
$$ 
We thus determine the optimal cut-off value $c_0$ as the one minimizing the distance between the point $(0, 1)$ and the ROC  curve, i.e., 
$$
    \widehat c_0 = \text{argmin}_{c}\sqrt{\big\{1 - \widehat{\text{TPR}}(c)\big\}^2 + \big\{1 - \widehat{\text{TNR}}(c)\big\}^2}.
$$
Finally, with the selected $\widehat{c}_0$, the estimated regression coefficient for an active index $\vi$ is given by
$$
    \widehat\vbeta_{\vi} = \frac{1}{I-B}\sum_{l=1}^{I-B} \valpha_{\vi}^{(B+l)}\cdot t(\valpha_{\vi}^{(B+l)};\lambda^{(B+l)}), ~~ \vi \in \widehat{\sV},
$$
and the corresponding estimated additive component function turns to be $\widehat f_{\vi} = \vphi\trans\widehat\vbeta_{\vi}$. 

\subsection{The Selection of Approximation Parameter}\label{supp:para_indi}
There are two considerations about the tuning parameter $\epsilon_0$ in the smooth indicator $t(\valpha_{\vi};\lambda)$ \eqref{smoo_indi}. On one hand, as required by MALA, the indicator should be smooth enough. On the other hand, as an indicator function, its range $[\min_{\vi}t(\valpha_{\vi};\lambda), \max_{\vi}t(\valpha_{\vi};\lambda)]$ needs to cover $[0,1]$ as much as possible.
\begin{figure}[htbp] 
		\centering 
		\includegraphics[width = 0.75\textwidth]{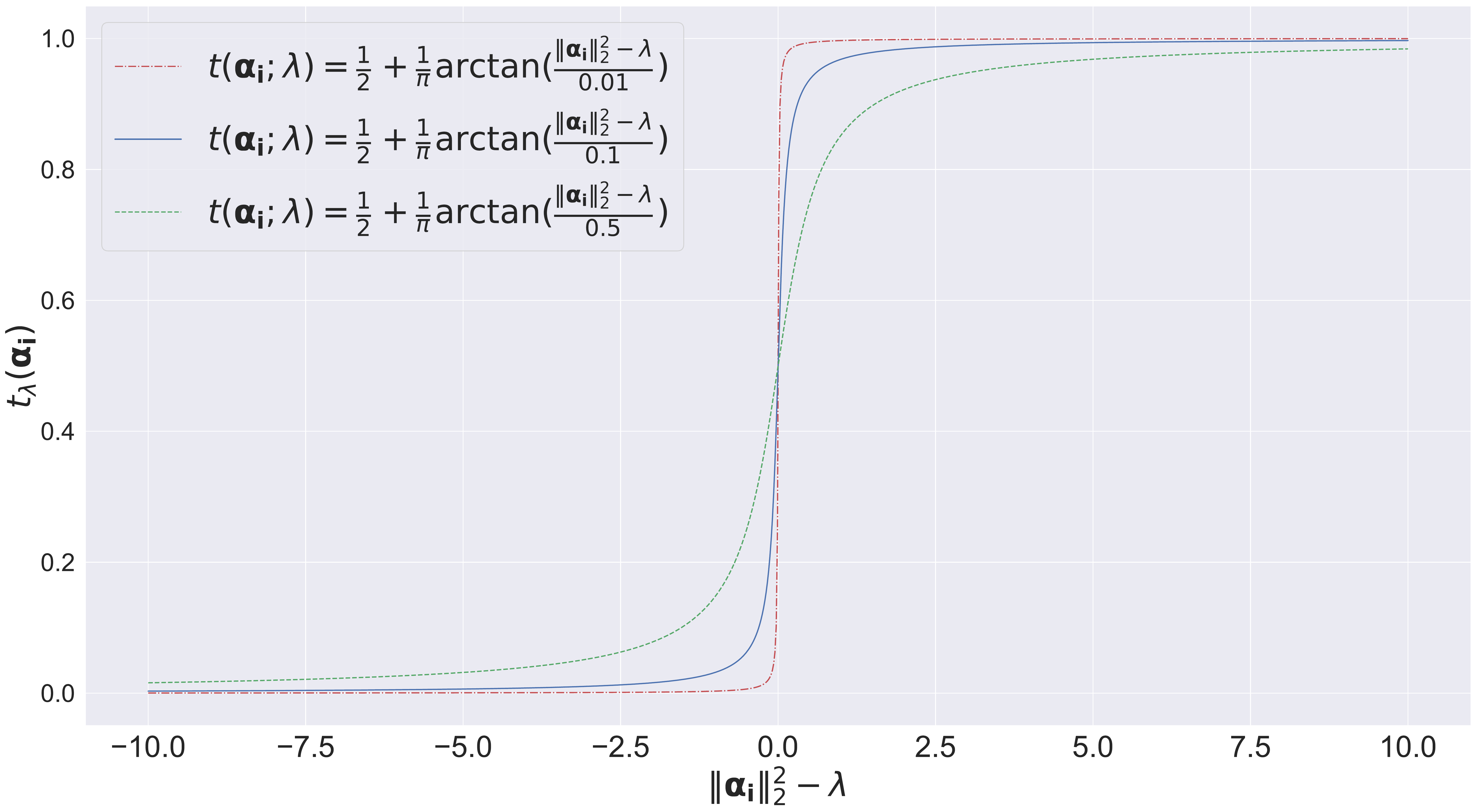} 
		\caption{The smooth indicators with different $\epsilon_0$'s.}
		\label{Indi}
\end{figure} 
According to Figure \ref{Indi}, we can see that with a bigger $\epsilon_0,$ the indicator becomes smoother, but $[\min_{\vi}t(\valpha_{\vi};\lambda), \max_{\vi}t(\valpha_{\vi};\lambda)]$ is harder to cover $[0,1]$. It should be avoided that the parameter $\epsilon_0$ is either too small or too big. Denote $m$ such that ${1}/{2} + ({1}/{\pi}) \arctan(m) = 1-\eta$, where $\eta$ is close to $0$. 
	To make $[\min_{\vi}t(\valpha_{\vi};\lambda), \max_{\vi}t(\valpha_{\vi};\lambda)]$ cover $[\eta, 1-\eta]$, we require $\epsilon_0$ to satisfy $\min_{\vi} \Vert\valpha_{\vi}\Vert_2^2 - \lambda \leq -m\epsilon_0$ and $\max_{\vi} \Vert\valpha_{\vi}\Vert_2^2 - \lambda \geq m\epsilon_0$, hence we have 
	$$\epsilon_0\in \left(0, \frac{\max_{\vi} \Vert\valpha_{\vi}\Vert_2^2 - \min_{\vi} \Vert\valpha_{\vi}\Vert_2^2}{2m}\right].$$ 
	We choose the largest value 
	$$\frac{\max_{\vi} \Vert\valpha_{\vi}\Vert_2^2 - \min_{\vi} \Vert\valpha_{\vi}\Vert_2^2}{2m}$$ 
	for $\epsilon_0$ to make the indicator smooth as far as possible, and we suggest to set $\eta$ as $0.05$ in the numerical experiments.

	However, in a real application, $\Vert\valpha_{\vi}\Vert_2^2$ is unknown. In practice, we apply a two-step strategy as follows to settle this problem. 
	\begin{enumerate}
		\item[(i)] 
			Set $t(\valpha_{\vi};\lambda)\equiv 1$ and drop the step updating $\lambda$ in Algorithm~~\ref{supp:MALA}. We run Algorithm~~\ref{supp:MALA} under $r = 1$ and $\rho_{\valpha} = \rho_{1}$ to get rough estimates of $\max_{\vi}\Vert\valpha_{\vi}\Vert_2^2$ and $\min_{\vi}\Vert\valpha_{\vi}\Vert_2^2$. We then set $$\epsilon_0 = \frac{\max_{\vi}\Vert\widehat\valpha_{\vi}\Vert_2^2 - \min_{\vi}\Vert\widehat\valpha_{\vi}\Vert_2^2}{2m}.$$ 
			\item[(ii)] With this $\epsilon_0$, we completely run Algorithm~~\ref{supp:MALA} under $r = 1$ and $\rho_{\valpha} = \rho_{1}$ to get new estimates of $\max_{\vi}\Vert\valpha_{\vi}\Vert_2^2$ and $\min_{\vi}\Vert\valpha_{\vi}\Vert_2^2$. We then set $$\epsilon_0 = \frac{\max_{\vi}\Vert\widehat\valpha_{\vi}\Vert_2^2 - \min_{\vi}\Vert\widehat\valpha_{\vi}\Vert_2^2}{2m}.$$
	\end{enumerate}	

  As for the tuning parameter $\epsilon_1$, it is used in the approximation to the $\ell_1$ function (see (\ref{smoo_fus}) of the main paper).
To closely approximate the nonsmooth $\ell_1$ function, $\epsilon_1$ is suggested to be small enough and we specify its value to be $10^{-6}$ as discussed in Section \ref{MALA_Warmstart} of the main paper. 
We find that our proposed model is not sensitive to the specific  value of $\epsilon_1$ through a sensitivity analysis on the simulated data.
In particular, we generated the simulated dataset of sample size 600 under the nonlinear setting of a horse shape with high SNR (Setting 5) as in Section 4 of the main paper.
On the simulated dataset, we applied our proposed BFEN method with $\epsilon_1 = 10^{-6}$, $10^{-8}$, and $10^{-10}$.
We repeated the experiments 30 times and calculated relative prediction error (RPE), mean squared error (MSE), relative mean squared error (RMSE), true positive rate (TPR), and true negative rate (TNR) as in the main paper. The results for various $\epsilon_1$'s are summarized in Table \ref{table_sensitivity}.
Table \ref{table_sensitivity} shows that our model is not sensitive to the tuning parameter $\epsilon_1$ with small values.

\begin{table}[h!]
	\centering
	
	\caption{ Estimation errors for different specifications of parameter $\epsilon_1$. The first row stands for the default specification, i.e., $\epsilon_1 = 10^{-6}$. The following rows summarize results when $\epsilon_1$ is assigned new values. The numbers in parentheses are the standard errors based on 30 replicates. 
	}
	\resizebox{\textwidth}{!}{
    	\begin{tabular}{ l| c c c c c}
		\hline
		\multicolumn{1}{c}{} & RPE & MSE & RMSE & TPR & TNR\\
			\hline
			\hline 
		$\epsilon_1 = 10^{-6}$ & $0.0763$ $(0.0024)$ &$0.0166$ $(0.0006)$ & $0.0508$ $(0.0024)$ & $0.9996$ $(0.0004)$  &$0.9392$ $(0.0021)$\\
            $\epsilon_1 = 10^{-8}$ &$0.0761$ $(0.0022)$ &$0.0165$ $(0.0007)$ & $0.0509$ $(0.0025)$ & $0.9998$ $(0.0002)$ &$0.9405$ $(0.0017)$ \\
            $\epsilon_1 = 10^{-10}$ &$0.0763$ $(0.0022)$ &$0.0168$ $(0.0008)$ & $0.0516$ $(0.0028)$ & $0.9996$ $(0.0004)$ & $0.9412$ $(0.0017)$ \\
            \hline 
	\end{tabular}		} 
	\label{table_sensitivity}
\end{table}		
	
\subsection{Validation Method}\label{supp:validation}
    \LinesNumberedHidden
	\begin{algorithm}[htpb]
		\caption{Validation method to select $p_0$ and $\rho_{\valpha}$.}
		\label{supp:Warmstart_1}
		Let $\text{Loss}_{j, t}$ denote the validation loss corresponding to hyperparameters $p_0 = p_{0,j}$ and $\rho_{\valpha} = \rho_{\valpha,t}$.
		
		Denote $\mTheta^{(j, t, i)} = \{\mu^{(j, t, i)}, \valpha^{(j, t, i)}, (\sigma^2)^{(j, t, i)}, \delta^{(j, t, i)}, \lambda^{(j, t, i)}\}$ the $i$-th update of posterior samples set under $p_0 = p_{0,j}$ and $\rho_{\valpha} = \rho_{\valpha, t}$.
		
		Set $p_{0,1} < \cdots < p_{0,J}, \rho_{\valpha,1}< \cdots < \rho_{\valpha,T},$ $W < B  < I$.
		
		Set $p_0 = p_{0,1}, \rho_{\valpha} = \rho_{\valpha,1},$ and initialize the parameter set $\mTheta^{(1, 1, 1)}$.

		Obtain $\{\mTheta^{(1, 1, i)}\}_{i = 1}^I$ through Algorithm \ref{supp:MALA} and set $j = 1, t = 2, j' = 1, t' = 1$.

		\While{$j \leq J$}{
			Set $r = r_j, \rho_{\valpha} = \rho_{\valpha,t}.$
			
			Initialize the parameters set $\mTheta^{(j, t, W)}$ with the averaged $\{\mTheta^{(j', t', i)}\}_{i = B}^I$.
			
			Obtain $\{\mTheta^{(j, t, i)}\}_{i = W}^I$ through Algorithm \ref{supp:MALA}, and compute validation loss $\text{Loss}_{j, t}.$
			
			Set $j' = j, t' = t$.
			
			\eIf{$j\equiv 1 \pmod{2}$}
			{\eIf{$t<T$}{$t = t + 1;$}
				{$j = j + 1$}
			}{\eIf{$t>1$}{$t = t - 1;$}
				{$j = j + 1$}	
			}
		}
		Set $(j_0, t_0) = \argmin_{j, t}\{\text{Loss}_{j, t}\mid 1\leq j\leq J, 1\leq t\leq T\}.$
		
	\end{algorithm}

	We apply the validation method to select tuning parameters $(r, \rho_{\valpha})$ in prior (\ref{naive_p}) and $p_0$ in hyperprior \eqref{prior_del_sig} from a corresponding list of candidate values  $\{r_s\}_{s = 1,\cdots,S}$, $\{\rho_{\valpha,t}\}_{t = 1,\cdots, T}$, and $\{p_{0,j}\}_{j = 1,\cdots,J}$. Given the estimated  $\{\widehat f_{\vi}\}_{\vi\in\sI}$ from the training set under $p_0 = p_{0,j}$, $r = r_s$, $\rho_{\valpha} = \rho_{\valpha, t}$,  the response value  $y$ is predicted in the validation set. The final tuning parameters are selected as those minimizing the validation loss $\text{L}(\vy_{\text{valid}}, \widehat \vy_{\text{valid}})$.
	
	For the validation method, applying Algorithm~\ref{supp:MALA} for all  combinations of $p_0\in\{p_{0,j}\}_{j = 1,\cdots,J}, r\in\{r_s\}_{s = 1,\cdots,S}, \rho_{\valpha}\in\{\rho_{\valpha,t}\}_{t = 1,\cdots, T}$ is a time-consuming process. We adopt the following strategy to reduce the computational cost. 
	
	\begin{enumerate}
		\item[(a)] The number of combinations of $(p_0, r,\rho_{\valpha})$ can be reduced from $J\cdot S \cdot T$ to $J\cdot T + S\cdot T$ by using a two-step greedy search:
		\begin{enumerate}
			\item[i)] Compute the validation loss under different $p_0\in\{p_{0,j}\}_{j = 1,\cdots,J}, \rho_{\valpha}\in\{\rho_{\valpha,t}\}_{t = 1,\cdots, T}$ with $r = r_1$ fixed.  Select $p_0 = p_{0,j_0}$ from the optimal pair $(p_0,\rho_{\valpha})$ that minimizes the  validation loss. 
			\item[ii)] Compute the validation loss under different $r\in\{r_s\}_{s = 1,\cdots,S}, \rho_{\valpha}\in\{\rho_{\valpha,t}\}_{t = 1,\cdots, T}$ with $p_0 = p_{0,j_0}$ fixed, then select the optimal $r = r_{s_0}, \rho_{\valpha,t_0}$. 	
		\end{enumerate}

		\item[(b)] The number of iterations in executing Algorithm~\ref{supp:MALA} under each $p_0,r,\rho_{\valpha}$ can be reduced by applying a warmstart. 
		In other words, the initial point of Algorithm~\ref{supp:MALA} under $r = r_2, \rho_{\valpha} = \rho_{\valpha,1}$ is determined as the output of Algorithm~\ref{supp:MALA} under $r = r_1, \rho_{\valpha} = \rho_{\valpha,1}$.
		We find that this initialization trick circularly reduces the computational burden of validation. 
	\end{enumerate}
The validation method to obtain the optimal tuning parameters $p_{0}$ and $\rho_{\valpha}$ are summarized in Algorithm~\ref{supp:Warmstart_1}. 
We omit the detailed algorithm to obtain the optimal tuning parameters $r$ since the procedure is similar.
For the candidate grids of $p_0$, $r$, and $\rho_{\valpha}$, their ranges should be reasonable and wide enough. 
Among them, the range of $r$ is within $[0, 1]$ and thus we assign the grid of $r$ as $\{1, 0.75, 0.5, 0.25, 0\}$ following \cite{zhou2020broadcasted}. 
For $\rho_{\valpha}$, we specify the grid $\{0.001, 0.005, 0.01, 0.05, 0.1, 0.5, 1, 5\}$ for $\rho_{\valpha}$ following the suggestion of \cite{teipel2015relative, engebretsen2019statistical,tec2020}. 
For $p_0$, its grid is suggested as
$$\{ 0.5 P_1\cdots P_D K,\, 5P_1\cdots P_D K,\, 50P_1\cdots P_D K, \, 500P_1\cdots P_D K,\, 5000P_1\cdots P_D K\},$$ 
where the lower bound of the grid is determined according to Proposition 1 of the main paper. 
We find all the above grids are wide enough in our numerical experiments.
\subsection{Sensitivity Analyses}\label{sens_ana}
\begin{figure}[h!] 
	\centering
	\includegraphics[width=0.8\textwidth]{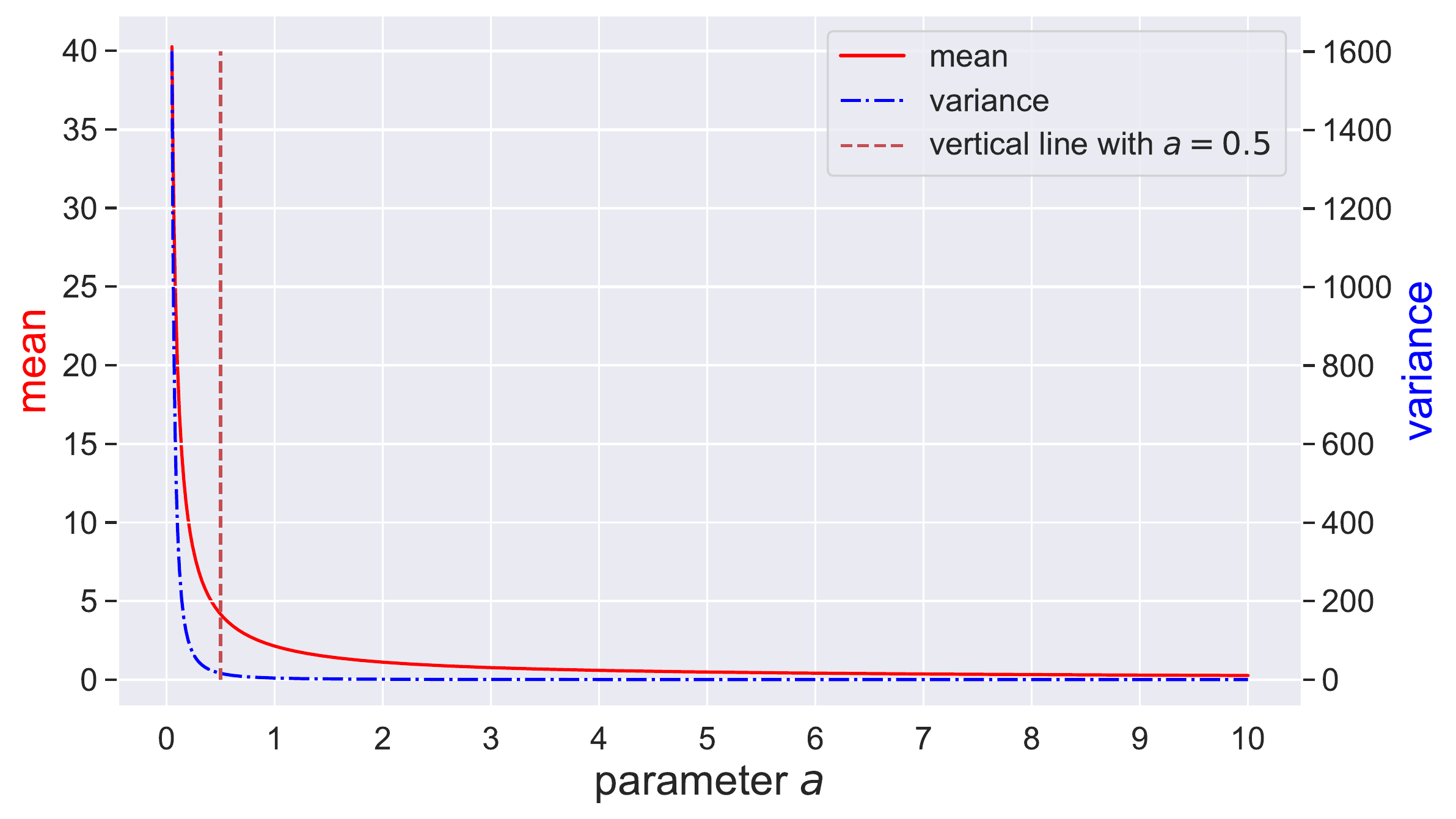} 
	\caption{The mean and the variance of the generalized inverse Gaussian prior (\ref{prior_mu_lambda}) as a function of parameter $a$ with $p$ and $b$ fixed as 1 and 0.5, respectively. 
 The plot has a shared $x$-axis and two $y$-axes correspondingly for mean (the left) and variance (the right).}
	\label{elbow}
	\end{figure}

For the other hyperparameters $\{p_1, \sigma^2_{\mu}, p, a, b\}$ and tuning parameters $\delta'$,
first note that according to Proposition~\ref{cons_thm} of the main paper, 
we set $p_1=(2P_1\cdots P_D - 1)K/2$ to balance the magnitude of the normalization term of the prior distribution $p(\delta,\sigma^2)$ in (\ref{prior_del_sig}) and thus $p(\delta,\sigma^2)$ is weakly informative with respect to $\sigma^2$ (similar to an inverse-gamma prior with the small shape parameter).
For the rest $\sigma^2_{\mu}$, $p$, $a$, $b$, and $\delta'$, we carry out a sensitivity analysis for these parameters. Recall that our default specifications are 
$\sigma^2_{\mu} = 100$, $p = 1$, $a = b = 0.5$, and $\delta' = 0.0001$, which renders the prior relatively weak-informative. 
In particular for $a$, 
we plot the mean and the variance as two functions of the parameter $a$ with $p$ and $b$ fixed at 1 and 0.5, respectively. According to Figure~\ref{elbow}, $a = 0.5$ shows reasonably weak-informative since it is at the ``elbow'' for both curves. 
In the sensitivity analysis, we consider a larger and a smaller values relative to the default setting for hyperparamters $\sigma^2_{\mu}$, $p$, $a$, $b$, and $\delta'$ to assess their sensitivity.  
Results of sensitivity analysis in the nonlinear setting of a horse shape with high SNR (Setting 5) of our simulation experiments are summarized in Table~\ref{table_sensitivity2} based on 30 replicates.
It can be seen that our model is relatively robust with different choices of tuning/hyper parameters. 

\begin{table}[h!]
 
	\centering
	\caption{Estimation errors for different specifications of tuning/hyper parameters. The first row shows the results with default specification: 
	$\sigma^2_{\mu} = 100$, $p = 1$, $a = b = 0.5$, and $\delta' = 0.0001$. The following rows exhibit the results with each parameter being assigned new values. The numbers in parentheses are the standard errors based on 30 replicates. 
	}
	\resizebox{\textwidth}{!}{
    	\begin{tabular}{ l| c c c c c}
		\hline
		\multicolumn{1}{c}{}& RPE & MSE & RMSE & TPR & TNR\\
			\hline
			\hline 
			default & $0.0763$ $(0.0024)$ &$0.0166$ $(0.0006)$ &$0.0508$ $(0.0024)$ & $0.9996$ $(0.0004)$  &$0.9392$ $(0.0021)$\\
			\hline
			\hline       
            $\sigma_{\mu}^2 = 10$ & $0.0751$ $(0.0021)$ & $0.0162$ $(0.0006)$ & $0.0493$ $(0.0017)$ & $1.0000$ $(0.0000)$ & $0.9404$ $(0.0024)$\\
            $\sigma_{\mu}^2 = 1000$ &$0.0748$ $(0.0019)$ &$0.0165$ $(0.0007)$ &$0.0505$ $(0.0026)$ &$0.9998$ $(0.0002)$ &$0.9405$ $(0.0019)$ \\
			\hline
			\hline    
            $p = - 10$ & $0.0765$ $(0.0024)$ &$0.0169$ $(0.0008)$ & $0.0521$ $(0.0028)$& $0.9995$ $(0.0004)$&$0.9404$ $(0.0019)$ \\
            $p = 10$ &$0.0757$ $(0.0020)$ &$0.0166$ $(0.0007)$ &$0.0514$ $(0.0022)$ & $1.0000$ $(0.0000)$& $0.9427$ $(0.0023)$\\        \hline
			\hline   
               $a = 0.25$ &${0.1002}$ $(0.0026)$&${0.0235}$ $(0.0016)$ &${0.0701}$ $(0.0048)$&$0.9991$ $(0.0004)$ & ${0.9237}$ $(0.0050)$\\
            $a = 1$ &${0.0667}$ $(0.0013)$ & ${0.0138}$ $(0.0003)$ &${0.0434}$ $(0.0011)$ &$1.0000$ $(0.0000)$ &${0.9487}$ $(0.0014)$ \\ \hline
			\hline    
            $b = 0.25$ &${0.0754}$ $(0.0021)$&${0.0165}$ $(0.0007)$ &${0.0489}$ $(0.0017)$&$1.0000$ $(0.0000)$ & ${0.9381}$ $(0.0025)$\\
            $b = 1$ &${0.0751}$ $(0.0020)$ & ${0.0164}$ $(0.0007)$ &${0.0488}$ $(0.0017)$ &$1.0000$ $(0.0000)$ &${0.9381}$ $(0.0025)$ \\ \hline
			\hline    
             $\delta' = 0.001$ & $0.0756$ $(0.0020)$ & $0.0163$ $(0.0006)$ & $0.0484$ $(0.0014)$ & $1.0000$ $(0.0000)$ & $0.9375$ $(0.0027)$ \\
            $\delta' = 0.00001$ & $0.0772$ $(0.0023)$ & $0.0167$ $(0.0007)$ & $0.0514$ $(0.0020)$ & $1.0000$ $(0.0000)$ & $0.9420$ $(0.0025)$ \\ \hline
			\hline      
	\end{tabular}		} 
	\label{table_sensitivity2}
\end{table}		

\subsection{Summary}\label{supp:summary_method}
	We summarize how to estimate the tensor regression model (\ref{eqn:model}) of the main paper. First, select the hyperparamters $\epsilon_0$ through the methods introduced in Section~\ref{supp:para_indi}. Second, follow Section~\ref{supp:validation}
	to obtain the optimal tuning parameters $(p_0,r, \rho_{\valpha})$. 
	Finally, follow Section~\ref{supp:Posterior} to obtain the estimated component functions $f_{\vi}$'s through the posterior samples corresponding to the optimal tuning parameters $(p_0, r, \rho_{\valpha})$. 
	
\section{Tuning Parameter Selection for the Compared Methods}\label{supp:para_exp}

We present here the details of tuning parameter selection for the competitive methods \citep{zhou2013tensor,hao2021sparse,BTR2017} in our numerical experiments.
As suggested in \cite{zhou2013tensor} and \cite{hao2021sparse}, we choose lasso penalty and group lasso penalty for FTR and STAR respectively, and also apply the validation method to select the rank of CP decomposition and the tuning parameters of their penalties. 
The training set, validation set, and test set for STAR and FTR are the same as those for BFEN. Note that BTR only needs a training set and a test set because it automatically selects the tuning parameters \citep{BTR2017}. So the training set and validation set for BFEN are  used together to train BTR.
	
For FTR and STAR, both of the tuning parameters of lasso (FTR) and group lasso (STAR) are selected from a geometric sequence $(\rho_1,\cdots, \rho_{10})$, where 
$\rho_1 = 0.1$ and $\rho_{10} = 10$; 
and the rank of CP decomposition is selected from $\{2, 4, 6, 8, 10\}$. 
The above grids are wide enough for two competing models in the sense that the boundary points are seldomly selected by either method.
For BTR, the hyperparamters are selected as \cite{BTR2017} suggested. In particular, the rank of the CP decomposition is selected as $10$  in simulation experiments. In real data experiments, we find that BTR fails to converge for some experiments among the $100$ replicates if the rank of CP decomposition is over $6$. Hence, we select the  rank as $5$ in real data experiments.

\section{Additional Results for the Simulation Study}

In this section we explain in details how to construct spatially smooth signals through graph Laplacian matrix, 
 and provide some additional outputs for the experimental results.

\subsection{The Construction of Spatially Smooth Model}\label{supp:setting_simu}

The parameters $\{\overline{u}_{ij}^{(1)}\}$, $\{v_{ij}^{(2)}\}$, and $\{v_{ij}^{(3)}\}$ in Table~\ref{table1} of the main paper are constructed through graph Laplacian matrix as follows.

First, we obtain the graph Laplacian matrix \citep{merris1994laplacian}, $\mL\in \mathbb{R}^{P_1P_2\times P_1P_2}$, of the graph $\sG =(\sI, \sE)$ defined in Section~\ref{Tensor_Additive} of the main paper where the index in $\sI$ is arranged in the column-major order.
Denote $\vu_l$ as the eigenvector of $\mL$ corresponding to the $l$-th smallest eigenvalue. 
We focus on the first $L$ ($L< P_1P_2$) eigenpairs with $l=1, \dots, L$. 
The eigenvector $\vu_l\in\bbR^{P_1P_2}$ is reshaped to be a matrix  $\mU_l\in\bbR^{P_1\times P_2}$ by the column-major order. 
As the eigenvector correspond to small Laplacian value, the element values of $\mU_l$ ($1\leq l \leq L$) has variability of low frequency and thus is spatially smooth across $(i,j)$ \citep{dong2016learning}. 
	
	Second, to make use of all the $L$ spatially smooth matrices, we further construct three matrices $\overline{\mU}^{(1)}$, $\overline{\mU}^{(2)}$, and $\overline{\mU}^{(3)}$ as random linear combinations of $\mU_1,\cdots, \mU_L$, i.e., 
	$\overline{\mU}^{(m)} = \sum_{l=1}^L \gamma_l^{(m)}  \mU_l$ with $\gamma_l^{(m)}  \stackrel{\mathrm{i.i.d.}}{\sim} \text{Unif}(0,1), l=1, \dots, L, m=1,2,3$. After that, $a_{ij}$ in non-linear cases of the horse shape (Settings 4 and 5 in Table~\ref{table1} of the main paper) is set as $1\overline{u}_{ij}^{(1)}+2$ 
 , where $\overline{u}_{ij}^{(1)}$ is the $(i,j)$-th element of $\overline{\mU}^{(1)}$.

	As for $c_{ij}$ and $d_{ij}$ in the non-linear cases of a horse shape and a shape of handwritten six (Settings 4, 5, 7, and 8 in Table~\ref{table1} of the main paper), they are constructed from $\overline{\mU}^{(2)}$ and $\overline{\mU}^{(3)}$. To be specific, we rescale each element ${u}_{ij}^{(m)}$ of $\overline{\mU}^{(m)}$, $m=2$ and $3$, to get a new matrix $\big(v_{ij}^{(m)}\big)$ such that $\min_{(i,j)\in\sV} v_{ij}^{(m)} = \pi$ and $\max_{(i,j)\in\sV} v_{ij}^{(m)} = 1.5\pi$, where $\sV$ is the set of active pixels. 
	We set $c_{ij}$ and $d_{ij}$ as $v_{ij}^{(2)}$ and $v_{ij}^{(3)}$, respectively. The number $L$ of eigenvectors is set as $L=80$.

\subsection{ Additional Results}\label{supp:map_simu}
\begin{figure}[t] 
		\centering 
		\includegraphics[width=\textwidth]{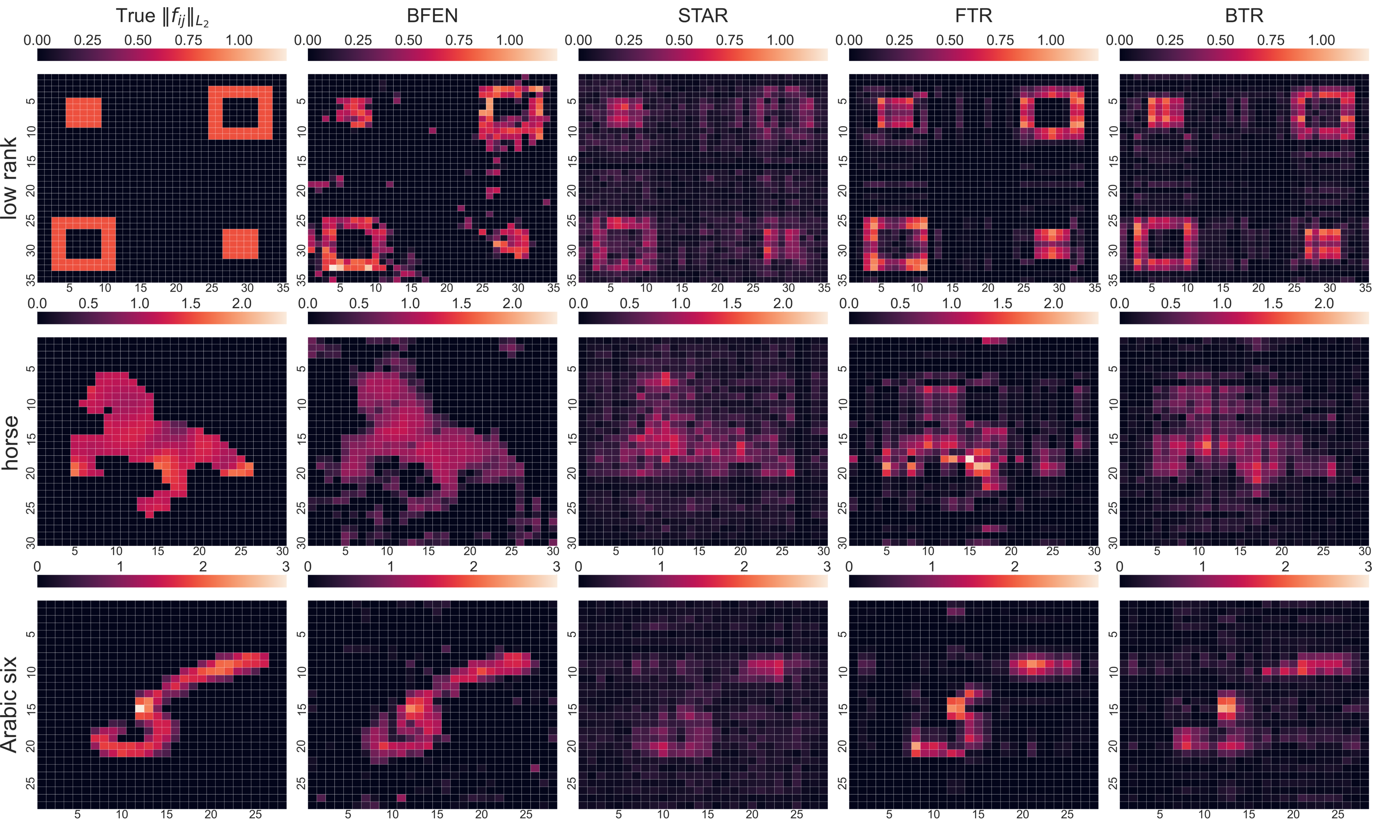} 
		\caption{ The heatmaps of various methods under the nonlinear with low SNR settings (Settings 1, 4, and 7).  Rows 1--3 correspond to the patterns of low-rank shapes (Setting 1), a horse shape (Setting 4), and a shape of handwritten Arabic six (Setting 7), respectively. The first column presents the magnitude of the true additive component function. Columns 2--5 correspond to the estimated results by BFEN, STAR, FTR, and BTR, respectively.}
		\label{supp:weak_simu}
	\end{figure}

\begin{figure}[t] 
		\centering 
		
		\includegraphics[width=\textwidth]{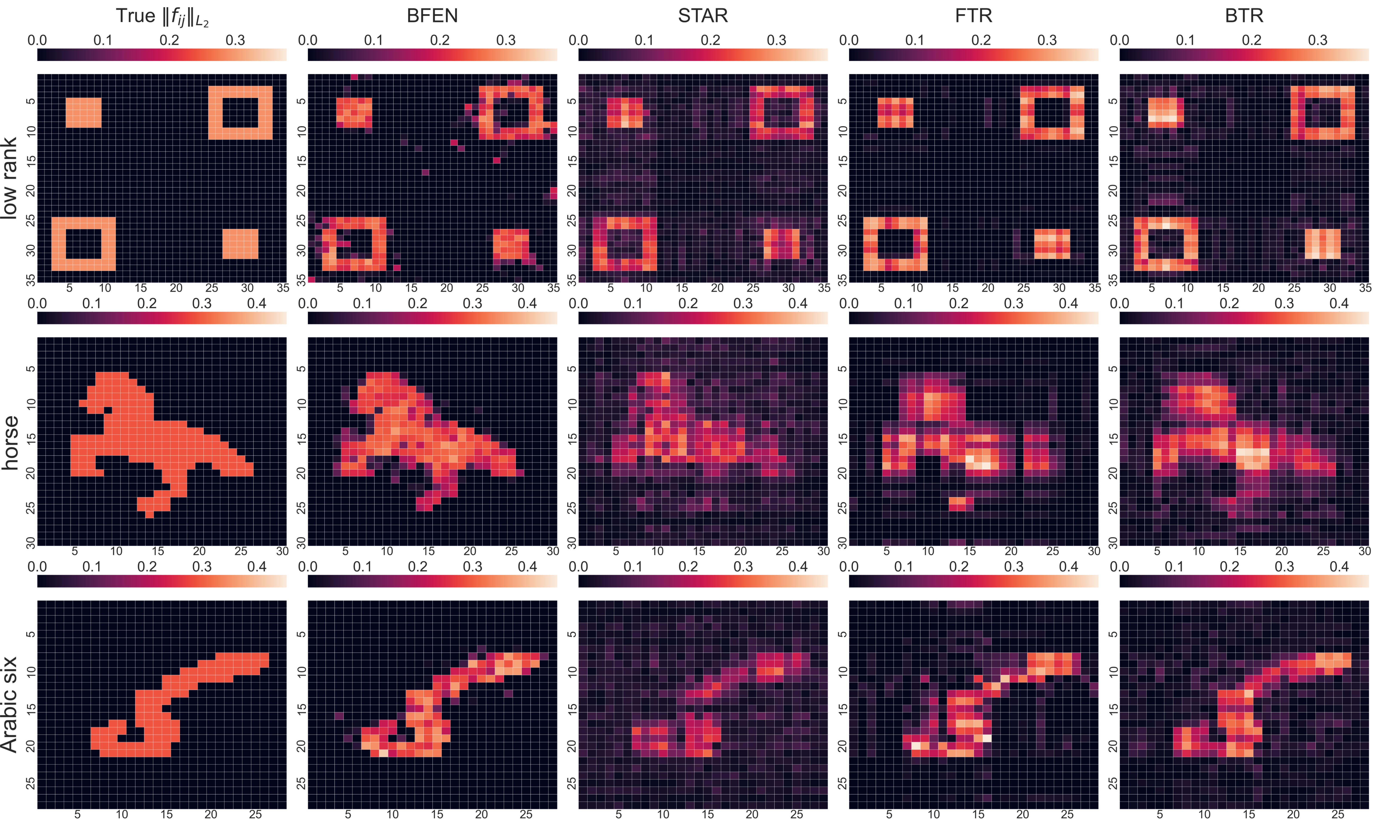} 
		\caption{ The heatmaps of various methods under the linear settings (Settings 3, 6, and 9).  Rows 1--3 correspond to the patterns of low-rank shapes (Setting 3), a horse shape (Setting 6), and a shape of handwritten Arabic six (Setting 9), respectively. The first column presents the magnitude of the true additive component function. Columns 2--5 correspond to the estimated results by BFEN, STAR, FTR, and BTR, respectively.}
		\label{supp:linear_simu}
	\end{figure}

	Recall that we have 9 simulation settings with 30 replicates for each setting. 
 We
provide the detailed numerical results and runtime for the simulation experiments in
Table~\ref{supp:table_simu}. All
methods were implemented on the same platform with a 2.2-GHz Intel E5-2650 v4 CPU. 
The results of the experiments under the nonlinear with high SNR settings have already been exhibited in Section~\ref{simulation_result} of the main paper by heatmaps. 
In this section, we exhibit the results with the median relative prediction error (RPE) under the nonlinear with low SNR and linear settings.
In Figures~\ref{supp:weak_simu} and~\ref{supp:linear_simu}, the shade of each square $(i,j)$ indicates the $\bbL_2$ norm of $f_{ij}$ as is in Figure~\ref{strong_simu} of the main paper. The heatmaps in the first column exhibits the magnitude  $\Vert f_{ij}\Vert_{\bbL_2}$ of the true component function, while in Columns 2--5 exhibit the magnitude $\Vert\widehat f_{ij}\Vert_{\bbL_2}$ estimated by BFEN, STAR, FTR, and BTR, respectively. 
Rows 1--3 correspond to the patterns of low-rank shapes (Settings 1 and 3), a horse shape (Settings 4 and 6) and a shape of handwritten six (Settings 7 and 9), respectively. 	
Figures~\ref{supp:weak_simu} and \ref{supp:linear_simu} show that our method outperforms STAR, FTR, and BTR for irregular sparsity shapes, i.e. a horse and a handwritten Arabic six. 
Moreover, it is also exhibited that all the methods have good performances when signals are linear and the shape of active region is of low rank. 
These results are consistent with Figures 
\ref{Visual_simu} and \ref{strong_simu} of the main paper.

\begin{table}[h!]
 
	\centering
	\caption{Average RPE, MSE, RMSE, TPR, TNR, and execution time (in minutes) of various methods in the simulation study. The reported time is the total execution time divided by the number of candidate parameter values in the grid of each method. The numbers in the parentheses are the standard errors based on 30 replicates. The best performances are boldfaced.}
	\resizebox{\textwidth}{!}{
    	\begin{tabular}{ c| c c c|  c c c |  c c c}
			\hline
			Setting ID & 1 & 2  & 3 & 4 & 5   & 6 & 7 & 8 & 9\\
		\hline
		Shape & \multicolumn{3}{c|}{Low rank} & \multicolumn{3}{c|}{Horse} & \multicolumn{3}{c}{Handwritten Arabic six}\\
    		\hline 
    		SNR & $5$ & $50$ & $5$& $5$ & $50$ & $5$& $5$ & $50$ & $5$\\
      \hline 
        \multirow{2}{*}{Setting Meaning}& low SNR & high SNR & \multirow{2}{*}{linear} & low SNR & high SNR & \multirow{2}{*}{linear} & low SNR & high SNR & \multirow{2}{*}{linear} \\
                   & nonlinear &  nonlinear & & nonlinear &  nonlinear & & nonlinear &  nonlinear & \\
			\hline
			\hline 
			& \multicolumn{9}{c}{$\text{RPE}$ $(\times 10^{-2})$} \\
			\cline{2-10}  
			BFEN & $ 58.49 ( 1.45 ) $& $ 16.87 ( 1.27 ) $& $ 29.79 ( 1.07 ) $& $ \textbf{37.09} ( 1.01 ) $& $ \textbf{7.63} ( 0.24 ) $& $ \textbf{23.99} ( 0.58 ) $& $ \textbf{38.98} ( 1.30 ) $& $ \textbf{8.39} ( 1.23 ) $& $ \textbf{20.62} ( 0.35 ) $\\
			STAR& $ 54.41 ( 1.13 ) $& $ \textbf{12.25} ( 0.65 ) $& $ 36.61 ( 0.87 ) $& $ 63.52 ( 1.02 ) $& $ 40.38 ( 0.66 ) $& $ 48.10 ( 0.75 ) $& $ 62.75 ( 1.15 ) $& $ 36.10 ( 1.04 ) $& $ 45.65 ( 0.68 ) $\\
			FTR& $ \textbf{46.00} ( 0.86 ) $& $ 28.01 ( 0.49 ) $& $ \textbf{21.46} ( 0.39 ) $& $ 63.27 ( 0.72 ) $& $ 49.98 ( 0.69 ) $& $ 36.91 ( 0.48 ) $& $ 54.79 ( 0.87 ) $& $ 39.67 ( 0.70 ) $& $ 29.55 ( 0.48 ) $\\
			BTR& $ 46.74 ( 0.77 ) $& $ 30.01 ( 0.52 ) $& $ 23.76 ( 0.45 ) $& $ 57.96 ( 0.65 ) $& $ 45.36 ( 0.51 ) $& $ 33.41 ( 0.38 ) $& $ 56.19 ( 0.72 ) $& $ 41.75 ( 0.65 ) $& $ 31.26 ( 0.37 ) $
 \\
			\hline 
			\hline 
			& \multicolumn{9}{c}{$\text{MSE}$ $(\times 10^{-2})$} \\
			\cline{2-10}
			BFEN& $ 3.86 ( 0.10 ) $& $ 1.15 ( 0.10 ) $& $ 0.18 ( 0.01 ) $& $ \textbf{6.96} ( 0.27 ) $& $ \textbf{1.66} ( 0.06 ) $& $ \textbf{0.16} ( 0.01 ) $& $ \textbf{6.13} ( 0.28 ) $& $ \textbf{1.51} ( 0.29 ) $& $ \textbf{0.05} ( 0.00 ) $\\
			STAR & $ 3.51 ( 0.10 ) $& $ \textbf{0.80} ( 0.05 ) $& $ 0.26 ( 0.01 ) $& $ 15.82 ( 0.23 ) $& $ 10.92 ( 0.15 ) $& $ 0.66 ( 0.01 ) $& $ 12.75 ( 0.26 ) $& $ 7.87 ( 0.21 ) $& $ 0.35 ( 0.01 ) $
\\
			FTR & $ \textbf{2.80} ( 0.06 ) $& $ 2.09 ( 0.02 ) $& $ \textbf{0.06} ( 0.00 ) $& $ 15.91 ( 0.16 ) $& $ 13.86 ( 0.13 ) $& $ 0.43 ( 0.01 ) $& $ 10.82 ( 0.12 ) $& $ 8.93 ( 0.09 ) $& $ 0.16 ( 0.00 ) $\\
			BTR& $ \textbf{2.80} ( 0.03 ) $& $ 2.24 ( 0.02 ) $& $ 0.09 ( 0.00 ) $& $ 14.04 ( 0.09 ) $& $ 12.43 ( 0.08 ) $& $ 0.35 ( 0.00 ) $& $ 10.97 ( 0.11 ) $& $ 9.37 ( 0.07 ) $& $ 0.17 ( 0.00 ) $\\
			\hline 
			\hline 
			& \multicolumn{9}{c}{$\text{RMSE}$ $(\times 10^{-2})$} \\
			\cline{2-10}
			BFEN & $ 28.43 ( 0.87 ) $& $ 10.34 ( 0.78 ) $& $ 9.26 ( 0.92 ) $& $ \textbf{13.61} ( 0.79 ) $& $ \textbf{5.08} ( 0.24 ) $& $ \textbf{6.01} ( 0.32 ) $& $ \textbf{25.36} ( 0.76 ) $& $ \textbf{9.14} ( 1.38 ) $& $ \textbf{3.33} ( 0.14 ) $\\
			STAR& $ 27.71 ( 1.01 ) $& $ \textbf{5.51} ( 0.41 ) $& $ 12.32 ( 0.54 ) $& $ 39.58 ( 0.67 ) $& $ 28.51 ( 0.49 ) $& $ 23.69 ( 0.55 ) $& $ 39.02 ( 0.89 ) $& $ 25.45 ( 0.76 ) $& $ 19.18 ( 0.54 ) $
\\
			FTR& $ 29.78 ( 0.48 ) $& $ 24.21 ( 0.24 ) $& $ 3.79 ( 0.19 ) $& $ 47.70 ( 0.54 ) $& $ 42.41 ( 0.44 ) $& $ 16.03 ( 0.38 ) $& $ 45.49 ( 0.47 ) $& $ 39.05 ( 0.45 ) $& $ 8.82 ( 0.20 ) $ \\
			BTR& $ \textbf{27.27} ( 0.32 ) $& $ 23.22 ( 0.15 ) $& $ \textbf{2.99} ( 0.12 ) $& $ 41.67 ( 0.35 ) $& $ 37.19 ( 0.30 ) $& $ 11.12 ( 0.22 ) $& $ 42.17 ( 0.38 ) $& $ 36.60 ( 0.33 ) $& $ 7.90 ( 0.20 ) $
\\
			\hline 
			\hline 
			& \multicolumn{9}{c}{$\text{TPR}$} \\
			\cline{2-10}
			BFEN& $ 0.96 ( 0.01 ) $& $ \textbf{1.00} ( 0.00 ) $& $ 0.99 ( 0.00 ) $& $ \textbf{0.99} ( 0.00 ) $& $ \textbf{1.00} ( 0.00 ) $& $ \textbf{0.99} ( 0.00 ) $& $ \textbf{0.98} ( 0.00 ) $& $ \textbf{0.99} ( 0.00 ) $& $ \textbf{1.00} ( 0.00 ) $
\\
			STAR& $ \textbf{1.00} ( 0.00 ) $& $ 0.99 ( 0.01 ) $& $ 0.99 ( 0.01 ) $& $ 0.98 ( 0.01 ) $& $ 0.97 ( 0.01 ) $& $ \textbf{0.99} ( 0.01 ) $& $ \textbf{0.98}( 0.01 ) $& $ 0.98 ( 0.01 ) $& $ 0.99 ( 0.01 ) $
\\
			FTR & $ 0.96 ( 0.01 ) $& $ 0.99 ( 0.00 ) $& $ \textbf{1.00} ( 0.00 ) $& $ 0.87 ( 0.01 ) $& $ 0.93 ( 0.01 ) $& $ 0.97 ( 0.00 ) $& $ 0.86 ( 0.01 ) $& $ 0.92 ( 0.01 ) $& $ 0.99 ( 0.00 ) $ \\
			BTR & $ 0.97 ( 0.00 ) $& $ \textbf{1.00} ( 0.00 ) $& $ \textbf{1.00} ( 0.00 ) $& $ 0.62 ( 0.01 ) $& $ 0.78 ( 0.01 ) $& $ 0.93 ( 0.00 ) $& $ 0.68 ( 0.01 ) $& $ 0.83 ( 0.01 ) $& $ 0.99 ( 0.00 ) $\\
			\hline 
			\hline 
			& \multicolumn{9}{c}{$\text{TNR}$} \\
			\cline{2-10}
			BFEN & $ 0.78 ( 0.02 ) $& $ 0.91 ( 0.00 ) $& $ 0.91 ( 0.00 ) $& $ 0.81 ( 0.01 ) $& $ 0.94 ( 0.00 ) $& $ 0.93 ( 0.00 ) $& $ 0.85 ( 0.01 ) $& $ 0.93 ( 0.00 ) $& $ 0.94 ( 0.00 ) $ \\
			STAR
& $ 0.00 ( 0.00 ) $& $ 0.00 ( 0.00 ) $& $ 0.00 ( 0.00 ) $& $ 0.00 ( 0.00 ) $& $ 0.00 ( 0.00 ) $& $ 0.00 ( 0.00 ) $& $ 0.00 ( 0.00 ) $& $ 0.00 ( 0.00 ) $& $ 0.00 ( 0.00 ) $
\\
			FTR & $ 0.32 ( 0.02 ) $& $ 0.29 ( 0.02 ) $& $ 0.27 ( 0.02 ) $& $ 0.28 ( 0.03 ) $& $ 0.24 ( 0.03 ) $& $ 0.21 ( 0.02 ) $& $ 0.26 ( 0.02 ) $& $ 0.22 ( 0.02 ) $& $ 0.16 ( 0.02 ) $ \\
			BTR& $ \textbf{0.99} ( 0.00 ) $& $ \textbf{0.98} ( 0.00 ) $& $ \textbf{0.98} ( 0.00 ) $& $ \textbf{0.98} ( 0.00 ) $& $ \textbf{0.97} ( 0.00 ) $& $ \textbf{0.97} ( 0.00 ) $& $ \textbf{0.99} ( 0.00 ) $& $ \textbf{0.98} ( 0.00 ) $& $ \textbf{0.98} ( 0.00 ) $
\\
			\hline 
			\hline 
			& \multicolumn{9}{c}{execution time (in minutes)} \\
			\cline{2-10}
			BFEN & $ 1.35 ( 0.02 ) $& $ 1.33 ( 0.02 ) $& $ 1.36 ( 0.02 ) $& $ 1.20 ( 0.02 ) $& $ 1.20 ( 0.02 ) $& $ 1.19 ( 0.01 ) $& $ 1.19 ( 0.04 ) $& $ 1.13 ( 0.01 ) $& $ 1.10 ( 0.01 ) $ \\
			STAR & $ 0.57 ( 0.03 ) $& $ 0.86 ( 0.04 ) $& $ 0.78 ( 0.03 ) $& $ 0.46 ( 0.02 ) $& $ 0.57 ( 0.03 ) $& $ 0.59 ( 0.03 ) $& $ 0.39 ( 0.02 ) $& $ 0.49 ( 0.02 ) $& $ 0.49 ( 0.02 ) $
\\
			FTR & $ 0.15 ( 0.00 ) $& $ 0.13 ( 0.00 ) $& $ 0.11 ( 0.00 ) $& $ 0.13 ( 0.00 ) $& $ 0.11 ( 0.00 ) $& $ 0.10 ( 0.00 ) $& $ 0.11 ( 0.00 ) $& $ 0.10 ( 0.00 ) $& $ 0.09 ( 0.00 ) $\\
			BTR & $ 17.73 ( 0.12 ) $& $ 17.81 ( 0.15 ) $& $ 17.20 ( 0.27 ) $& $ 14.33 ( 0.11 ) $& $ 14.52 ( 0.14 ) $& $ 14.18 ( 0.02 ) $& $ 13.07 ( 0.01 ) $& $ 12.95 ( 0.04 ) $& $ 12.93 ( 0.04 ) $\\   
			\hline    
	\end{tabular}		} 
	\label{supp:table_simu}
\end{table}

\subsection{A Comparative Study with Random Walk Metropolis}\label{supp:compare_RW}
In Algorithm~\ref{supp:MALA}, a hybrid method (Metropolis-adjusted Langevin Algorithm, MALA) and the smoothing technique (Eqn.~(\ref{smoo_indi}) and Eqn.~(\ref{smoo_fus})) are employed for the update of $\valpha$. We chose the MALA over the random walk metropolis due to its computational efficiency. To illustrate this point, 
we compare MALA and the random walk metropolis under the nonlinear setting of handwritten Arabic six with high SNR (Setting 8) of our simulation experiments. We apply the proposed BFEN model with MALA (Algorithm~\ref{supp:MALA} of the Appendix) and the random walk metropolis (the corresponding MALA step is replaced by a random walk metropolis step in Algorithm~\ref{supp:MALA}) on the simulated dataset to sample the coefficients $\valpha$ of the unknown functions. Note that since the smoothing technique is no longer involved for the random walk metropolis, $\epsilon_0$ and $\epsilon_1$ are released in this method.
The other tuning/hyper parameters of random walk metropolis are set in the same way as the MALA method. 
In other words, the two methods use the same tuning/hyper parameters except for the extra smoothing parameters in the MALA algorithm. 
We set the lengths of Markov chains to 20,000 and 50,000 for MALA and random walk metropolis, respectively.
In this experiment, the acceptance rate of random walk proposal is around $0.45$. 

\begin{figure}[h!] 
	\centering
	\includegraphics[width=0.6\textwidth]{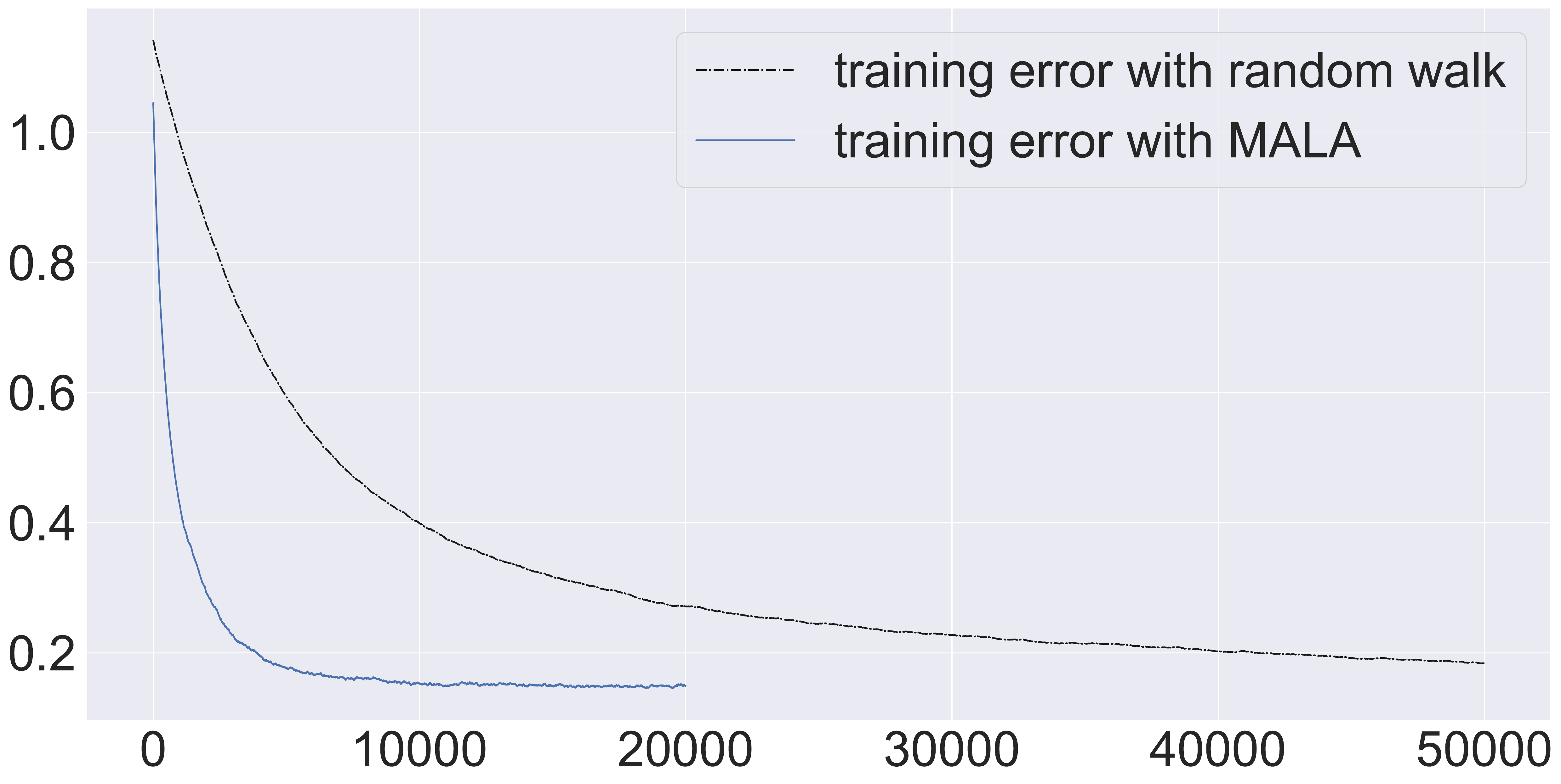} 
	\caption{ The trace plot of average training error $({1}/{N})\sum_{i = 1}^N(y_i - \widehat y_i)^2$ of the Markov chains for MALA and random walk metropolis for the first grid of tuning parameters (i.e., $r = 1$ and $\rho_{\valpha} = 0.001$). The plotted training error at each iteration is the average of 10 replicates. The slight difference of the initial training errors for two algorithms is caused by the extra smoothing approximation employed in MALA. 
    }
	\label{mala_rw}
	\end{figure}

   To inspect the convergence of MALA and random 
walk, we depict the trace plot of average training error $({1}/{N})\sum_{i = 1}^N(y_i - \widehat y_i)^2$, which is proportional to the negative log-likelihood, of both MALA and random walk metropolis.
The error is averaged over 10 replicates for 
the first candidate value of tuning parameters
in Figure~\ref{mala_rw}. 
Figure~\ref{mala_rw} reveals that random walk metropolis fails to explore the posterior efficiently, and that it has not yet converged even with a much longer Markov chain.

   We also calculate the relative prediction 
error (RPE), mean squared error (MSE), relative mean squared error (RMSE), true positive rate (TPR), and true negative rate (TNR). These results are based on the last 1,000 iterations of the two algorithms averaged over 10 replicates, which are summarized in Table~\ref{mala_rw_table}. The table also suggests the slow convergence of the random walk Metropolis algorithm. 

\begin{table}[H]
 
	\centering
	\caption{Operating characteristics for MALA and the random walk metropolis. The results are based on 10 replicates.  
	}
	\resizebox{0.8\textwidth}{!}{	\begin{tabular}{ l| c c c c c c}
	&  RPE & MSE & RMSE & TPR & TNR\\
		\hline

   MALA  &  $0.08$ $(0.02)$ & $0.01$ $(0.00)$& $0.09$ $(0.02)$ & $0.99$ $(0.00)$ & $0.93$ $(0.00)$\\
          \hline

   random walk &  $0.38$ $(0.03)$ & $0.09$ $(0.01)$& $0.31$ $(0.02)$ & $0.90$ $(0.01)$ & $0.90$ $(0.01)$ \\
\end{tabular}}
	\label{mala_rw_table}
\end{table}

\section{Additional Numerical Results for Facial Data Analysis}\label{supp:map_real}

\begin{figure}[!h] 
		\centering 
		\includegraphics[width=\textwidth]{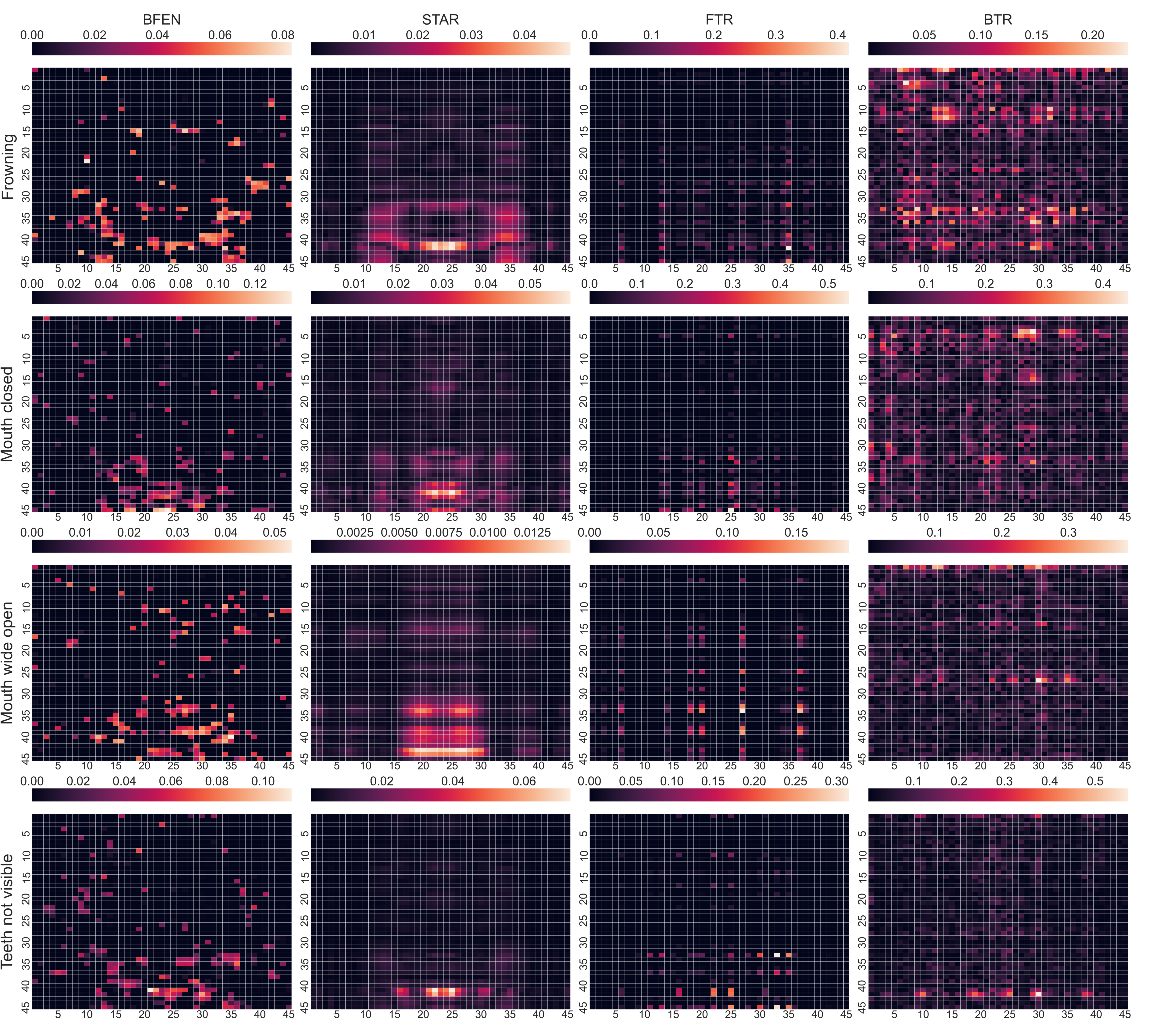} 
		\caption{The heatmaps for the response attributes {\it frowning}, {\it mouth closed}, {\it mouth wide open}, and {\it teeth not visible}. The shade of square $(i,j)$ in the heatmaps represents the $\bbL_2$ norm of $f_{ij}$. The heatmaps in Columns 1--4 correspond to the magnitude $\Vert\widehat f_{ij}\Vert_{\bbL_2}$ estimated by BFEN, STAR, FTR, and BTR, respectively.}
		\label{supp:real1}
	\end{figure}

 We provide the detailed numerical results and runtime for the facial data analysis in Table \ref{supp:table_simu}. All algorithms were run on the same platform with a 2.2-GHz Intel E5-2650 v4 CPU. 
The magnitude of each estimated additive component functions for the response attribute {\it smiling} have already been presented in Section~\ref{real_application} of main paper as heatmaps. In this section, we depict the heatmaps of other facial attribute {\it frowning}, {\it mouth closed}, {\it mouth wide open}, and {\it teeth not visible} in the different rows of Figure~\ref{supp:real1}. The selected heatmap corresponds to the replicate with the median RPE for each method. 
	
It is evident from the figure that BFEN has better interpretability in most cases. 
 Similar to the attribute {\it smiling}, the result of {\it frowning} in the first row of Figure~\ref{supp:real1} is also determined by the pixel values around the eyes, mouth and some facial muscles. The attribute {\it mouth closed} can be determined by  the positions of a person's lips and the skin around the lips. When someone keeps his/her {\it mouth wide open}, the upper lip and lower lip are apart, and thus the mouth cavity can be detected from the image. 
The teeth is obviously critical for the prediction of the attribute {\it teeth not visible}. Besides, part of the muscles (orbicularis oris) around the lips are also related to this attribute. 
In contrast, all the results of FTR and BTR lack  interpretations. With the help of nonlinearity and the group regularization across different blocks, STAR has better interpretability than FTR and BTR, but is still inferior to BFEN due to its low-rank modeling. For example, the rectangular subregion selected by STAR may not sufficiently interpret the attribute {\it mouth wide open}. Overall, our method can achieve a better balance between interpretation and predictive accuracy.

\begin{table}[h!]
  
	\caption{Average RPE and execution time (in minutes) of various methods for each attribute of the facial data analysis. The reported time is the total execution time divided by the number of candidate parameter values in the grid of each method. The numbers in the parentheses are the standard errors based on 100 replicates of random splitting. The best performances are boldfaced. }
	\centering
	\resizebox{\textwidth}{!}{
		\begin{tabular}{ c|c c c c c}
			\hline
				\
				Attribute & Smiling & Frowning & Mouth closed & Mouth wide open & Teeth not visible\\
				\hline 
				\hline 
				& \multicolumn{5}{c}{$\text{RPE}$} \\
								\cline{2-6}
				BFEN & $ \textbf{0.2129} $ $( 0.0015 ) $& $ \textbf{0.2198} $ $( 0.0013 ) $& $ \textbf{0.4510} $ $( 0.0025 ) $& $ \textbf{0.2365} $ $( 0.0013 ) $& $ \textbf{0.3209} $ $( 0.0019 ) $
				\\
				STAR & $ 0.2233 $ $( 0.0014 ) $& $ 0.2314 $ $( 0.0015 ) $& $ 0.4647 $ $( 0.0027 ) $& $ \textbf{0.2369} $ $( 0.0012 ) $& $ 0.3260 $ $( 0.0018 ) $ \\
				FTR & $ 0.2296 $ $( 0.0015 ) $& $ 0.2407 $ $( 0.0016 ) $& $ 0.5117 $ $( 0.0032 ) $& $ 0.2621 $ $( 0.0016 ) $& $ 0.3449 $ $( 0.0022 ) $
\\
				BTR & $ 0.2501 $ $( 0.0026 ) $& $ 0.2599 $ $( 0.0024 ) $& $ 0.5136 $ $( 0.0042 ) $& $ 0.2641 $ $( 0.0024 ) $& $ 0.3748 $ $( 0.0038 ) $ \\
                \hline 
				\hline 
				& \multicolumn{5}{c}{$\text{execution time $ $(in minutes)}$} \\
								\cline{2-6}
				BFEN & $ 1.9282 $ $( 0.0255 ) $& $ 1.7675 $ $( 0.0189 ) $& $ 1.9220 $ $( 0.0251 ) $& $ 1.9279 $ $( 0.0256 ) $& $ 1.9364 $ $( 0.0253 ) $
				\\
				STAR & $ 2.1964 $ $( 0.0254 ) $& $ 2.2022 $ $( 0.0241 ) $& $ 1.9066 $ $( 0.0265 ) $& $ 2.0286 $ $( 0.0400 ) $& $ 2.4710 $ $( 0.0686 ) $ \\
				FTR & $ 0.7650 $ $( 0.0099 ) $& $ 0.7563 $ $( 0.0101 ) $& $ 0.4835 $ $( 0.0135 ) $& $ 0.4515 $ $( 0.0116 ) $& $ 0.6747 $ $( 0.0111 ) $\\
				BTR & $ 25.4650 $ $( 0.2691 ) $& $ 27.1958 $ $( 0.2461 ) $& $ 26.2883 $ $( 0.2594 ) $& $ 28.3153 $ $( 0.3101 ) $& $ 26.6684 $ $( 0.3309 ) $
 \\                 
              \hline
				
		\end{tabular}
	}
	\label{supp:table_real_RPE}
\end{table}		
 
	\bibliographystyle{jasa} 
	\bibliography{citation}	
\end{document}